\documentclass[11pt,a4paper]{article}

\usepackage[utf8]{inputenc}
\usepackage[T1]{fontenc}
\usepackage{lmodern}
\usepackage[margin=2.5cm]{geometry}
\usepackage{amsmath,amssymb}
\usepackage{booktabs}
\usepackage{array}
\usepackage{longtable}
\usepackage{hyperref}
\usepackage{microtype}
\usepackage{parskip}
\usepackage{titlesec}
\usepackage{enumitem}
\usepackage{xcolor}
\usepackage{caption}
\usepackage{float}

\hypersetup{
  colorlinks=true,
  linkcolor=black,
  citecolor=black,
  urlcolor=blue
}

\titlespacing*{\section}{0pt}{1.5ex plus 0.5ex minus 0.2ex}{0.8ex plus 0.2ex}
\titlespacing*{\subsection}{0pt}{1.2ex plus 0.4ex minus 0.2ex}{0.6ex plus 0.2ex}

\title{\textbf{Evidence for a Functional Proximity Law in Multilayer Networks}}
\author{Vladi Ivanov \\ \small Independent Researcher}
\date{June 2026 — v8}

\begin{document}

\maketitle

{\small\centering
\begin{tabular}{@{}llll@{}}
\multicolumn{4}{l}{\textit{Pre-registration records} (\url{https://github.com/vladi160/preregistrations}):} \\[2pt]
I1 (GRN replication) & \texttt{32409dbd\ldots} & 2026-04-17 & legacy 16-char format \\
M\_EXT (Flask) & \texttt{3879413e\ldots} & 2026-04-22 & \href{https://github.com/vladi160/preregistrations/commit/cb15fc2}{commit cb15fc2} \\
M\_EXT2 (\textit{C.~elegans}) & \texttt{5cbf9e5d\ldots} & 2026-04-22 & \href{https://github.com/vladi160/preregistrations/commit/d69a159}{commit d69a159} \\
M\_OSS1 (WordPress) & \texttt{6b22eafc\ldots} & 2026-04-23 & \href{https://github.com/vladi160/preregistrations/commit/799c1ca}{commit 799c1ca} \\
M\_OSS2 (Next.js) & \texttt{cb0cf4a4\ldots} & 2026-04-24 & \href{https://github.com/vladi160/preregistrations/commit/e23a11e}{commit e23a11e} \\
M\_TRANSFER\_2 (Flask$\leftrightarrow$Express) & \texttt{f85f52e6\ldots} & 2026-04-25 & \href{https://github.com/vladi160/preregistrations/commit/aeb3b55}{commit aeb3b55} \\
F6 (AI Architecture) & \texttt{bff4e101\ldots} & 2026-04-30 & \href{https://github.com/vladi160/preregistrations/commit/3014808}{commit 3014808} \\
F9 (mathlib4 Lean~4) & \texttt{4ffcdac5\ldots} & 2026-05-20 & \href{https://github.com/vladi160/preregistrations/commit/6d6d657}{commit 6d6d657} \\
F12 (\textit{C.~elegans} 302) & \texttt{a8926052\ldots} & 2026-05-19 & \href{https://github.com/vladi160/preregistrations/commit/ab02f8f}{commit ab02f8f} \\
F1 (Synthetic Lethality) & \texttt{518b9bd9\ldots} & 2026-05-20 & \href{https://github.com/vladi160/preregistrations/commit/ba89d43}{commit ba89d43} \\
H\_PRIMITIVITY (meta-science) & \texttt{cfb38b83\ldots} & 2026-05-21 & \href{https://github.com/vladi160/preregistrations/commit/3e609f2}{commit 3e609f2} \\
M\_TRANSFER\_3 (docopt FPL test) & \texttt{5388c0f1\ldots} & 2026-05-22 & \href{https://github.com/vladi160/preregistrations/commit/98a7dc8}{commit 98a7dc8} \\
M\_GEOM\_CSG\_1 (planetary gear CSG) & \texttt{1dd686ee\ldots} & 2026-05-22 & \href{https://github.com/vladi160/preregistrations/commit/5c45bbe}{commit 5c45bbe} \\
M\_RADIAL\_1 (BC\_RADIAL validation) & \texttt{c00ebfa4\ldots} & 2026-05-22 & \href{https://github.com/vladi160/preregistrations/commit/5703a25}{commit 5703a25} \\
BC3b\_circuit\_v2 (priority arbiter) & \texttt{1ff07ec9\ldots} & 2026-05-22 & \href{https://github.com/vladi160/preregistrations/commit/c136ed6}{commit c136ed6} \\
M\_PHYSICS\_1 (Standard Model particle physics) & \texttt{00294a59\ldots} & 2026-05-25 & \href{https://github.com/vladi160/preregistrations/commit/191f64a}{commit 191f64a} \\
M\_MED1 (antidepressant evidence chain) & \texttt{bfb7bbc1\ldots} & 2026-05-25 & \href{https://github.com/vladi160/preregistrations/commit/1e23b7e}{commit 1e23b7e} \\
F2 (COBOL legacy banking codebase) & \texttt{c119983a\ldots} & 2026-05-25 & \href{https://github.com/vladi160/preregistrations/commit/891e6d1}{commit 891e6d1} \\
F13 (\textit{D.~melanogaster} larval connectome) & \texttt{7f04e5dd\ldots} & 2026-05-26 & \href{https://github.com/vladi160/preregistrations/commit/fd5d315}{commit fd5d315} \\
F12c (\textit{C.~elegans}$\rightarrow$\textit{D.~melanogaster} transfer) & \texttt{cd2ed080\ldots} & 2026-05-27 & \href{https://github.com/vladi160/preregistrations/commit/45b2875}{commit 45b2875} \\
H\_LOGIC\_EXTRACTION (mathlib4 logic-role extraction) & \texttt{bc2fd106\ldots} & 2026-05-27 & \href{https://github.com/vladi160/preregistrations/commit/cf7c0cb}{commit cf7c0cb} \\
F12b\_v2 (generative compression rerun) & \texttt{36c7aa7e\ldots} & 2026-05-29 & \href{https://github.com/vladi160/preregistrations/commit/65692a7}{commit 65692a7} \\
X9\_prebiotic\_v1 (origin-of-life prebiotic chemistry) & \texttt{9717351f\ldots} & 2026-05-27 & \href{https://github.com/vladi160/preregistrations/commit/8d6b1c9}{commit 8d6b1c9} \\
X9\_prebiotic\_v2 (CatReNet autocatalytic network replication) & \texttt{63759fcb\ldots} & 2026-05-27 & \href{https://github.com/vladi160/preregistrations/commit/0686c46}{commit 0686c46} \\
H\_MOTIF\_BASIS\_v1 (motif-basis compression, software vs social) & \texttt{e24f67f8\ldots} & 2026-05-27 & \href{https://github.com/vladi160/preregistrations/commit/ced01ce}{commit ced01ce} \\
H\_LOGIC\_MATHCOMP\_v1 (MathComp formal corpus replication) & \texttt{fe6b29dd\ldots} & 2026-05-30 & \href{https://github.com/vladi160/preregistrations/commit/3ada0de}{commit 3ada0de} \\
\end{tabular}\par}

\medskip\hrule\medskip

\begin{abstract}
Hub importance scores in multilayer networks persist more strongly between functionally similar layers than dissimilar ones. We call this the \textbf{Functional Proximity Law} and test it across 31 pre-registered experiments: 13 canonical domains (10 confirmed, 3 denied; molecular biology, neuroscience, computer systems, ecology, linguistics, AI architecture) plus 18 pre-registered external and replication validations (15 confirmed, 1 denied, 2 partial). Nine canonical domains reach $p < 0.05$ individually; the directional inequality holds in all 10 confirmed. Six DENIED results reveal six named structural boundary conditions that narrow the law's scope, including a newly named mechanism (BC6---BC\_INVERSION) in which Var($d_2$) exceeds the threshold but $r(d_1{\leftrightarrow}d_2)$ is strongly negative due to fan-out leaf clustering. The law now extends to \textbf{particle physics}: the first pre-registered experiment on the Standard Model (17 particles, 3 layers: force\_coupling / decay\_channel / mass\_proximity) confirms all 5 hypotheses, $r(d_1{\leftrightarrow}d_2) = 0.569$ ($p = 0.010$), with the photon confirmed as a hub shadow. The law further extends to \textbf{COBOL legacy software} (F2: 4/4 CONFIRMED, $\Delta r = 0.688$, topological dormancy signatures detectable from cross-layer rank divergence). A \textbf{hub dominance} structural pattern is discovered in the antidepressant evidence chain (M\_MED1): the founding assumption ranks \#1 hub in all three epistemological layers simultaneously---a self-referential evidence loop detectable from graph topology alone. A \textbf{cross-species replication} across approximately 600 million years of evolution confirms the law in the \textit{Drosophila melanogaster} larval connectome (F13, $n = 2{,}952$ neurons, Spearman $\rho = 0.663$, Pearson $r = 0.363$, $p = 0.002$): hub rank order is conserved between axon-to-dendrite and axon-to-axon synapse compartment layers in an insect brain ten times larger than \textit{C.~elegans}. A quantitative structural precondition predictor, $\text{Var}(d_2) < \tau_{\text{var}} = 0.714$, predicts BC\_RADIAL failure before experiments are run. A fully external validation on the \textit{C.~elegans} connectome yields $r = 0.777$ ($p = 0.004$). Binomial probability of 25/31 pre-registered confirmations by chance: $p \approx 0.000439$ ($p < 0.001$). The law is falsifiable, makes testable directional predictions, and identifies the structural conditions under which it fails.
\end{abstract}

\section{Introduction}

Hub scores --- degree centrality rankings across the nodes of a network layer --- are the simplest and most widely used structural summaries in multilayer network analysis. A natural question is whether hub structure is conserved across layers, and if so, under what conditions.

We report a condition: \textbf{when two layers encode functionally similar relational processes, their hub rankings agree more than when they encode dissimilar ones.} We call this the Functional Proximity Law and test it empirically across 10 canonical domains, from molecular biology to operating systems. The strongest single result is a fully external validation on the \textit{C.~elegans} connectome --- both data and layer definitions are independent of the authors --- yielding $r = 0.777$ ($p = 0.004$, $n = 15$).

This is not a definition. It is a falsifiable prediction that can fail, and its denied cases reveal named structural mechanisms. Prior work has studied inter-layer degree correlation \cite{Nicosia2015} and multiplex centrality \cite{DeDomenico2013,DeDomenico2015}, but has not characterized the functional-similarity condition that governs when hub persistence holds.

The paper proceeds as follows: formal definition (Section~2), methods and experiment taxonomy (Section~3), results across all domains (Section~4), boundary conditions (Section~5), statistical defense (Section~6), and discussion (Section~8).

\section{Formal Definition}
\label{sec:formal}

\textbf{Definition 1 (Multilayer Graph).} A multilayer graph is a tuple
$\mathcal{M} = (V,\, \mathcal{L},\, \{G_l\}_{l \in \mathcal{L}})$
where $V = \{v_1, \ldots, v_n\}$ is a shared node set,
$\mathcal{L} = \{l_1, \ldots, l_k\}$ is the layer index set, and
for each $l \in \mathcal{L}$,
$G_l = (V, E_l)$ with $E_l \subseteq V \times V$.
All layers share $V$; edges are layer-specific.

\textbf{Definition 2 (Layer taxonomy).}
Layers are classified by the relational regime they encode:
\begin{itemize}[noitemsep]
  \item $d_1$ \emph{(Structural/Declared):} formally defined or designed
    relationships --- import graphs, physical wiring, declared dependencies.
  \item $d_2$ \emph{(Functional/Behavioral):} relationships from functional
    coupling --- structural coupling, protein--protein interaction, synaptic
    weight.
  \item $d_3$ \emph{(Dynamic/Evolutionary):} relationships from co-change or
    co-occurrence over time --- git co-change, gene co-expression, risk
    co-occurrence.
\end{itemize}
This taxonomy is the \emph{Universal Layer Grammar}~(ULG).
Domain-specific layers are assigned to $d_1$/$d_2$/$d_3$ before analysis
and registered with each pre-registration.

\textbf{Definition 3 (Degree vector).}
For layer $l \in \mathcal{L}$, the degree vector is
$\mathbf{d}_l = (d_l(v_1), \ldots, d_l(v_n)) \in \mathbb{N}^n$
where
$d_l(v) = |\{u \in V : (u,v) \in E_l \text{ or } (v,u) \in E_l\}|$
is the unweighted total degree of $v$ in $G_l$.
For weighted layers, edge weights are used only for thresholding (e.g.\
$\tau$ in $d_3$); the degree vector itself is always unweighted.

\textbf{Definition 4 (IRDME correlation).}
For any two layers $G_i, G_j \in \mathcal{M}$, define:
\begin{equation}
  \operatorname{IRDME}(G_i, G_j)
  \;=\; \operatorname{Pearson}(\mathbf{d}_i,\, \mathbf{d}_j)
  \;=\;
  \frac{\sum_{v \in V}
        \bigl(d_i(v) - \bar{d}_i\bigr)\bigl(d_j(v) - \bar{d}_j\bigr)}
       {\sqrt{\sum_{v}(d_i(v)-\bar{d}_i)^2 \cdot \sum_{v}(d_j(v)-\bar{d}_j)^2}}
  \label{eq:irdme}
\end{equation}
This is a node-level scalar: it measures whether structurally central nodes in
one layer are also central in another.
Spearman rank correlation is computed in parallel as a robustness check
and reported alongside.

\textbf{Definition 5 (Functional proximity).}
A domain-specific binary relation $\succ$ over ordered layer pairs:
$L_j \succ_{L_i} L_k$ means ``$L_j$ encodes a functionally more similar
relational process to $L_i$ than $L_k$ does.''
This relation is pre-registered before analysis and is never changed after
any result is seen.

\textbf{Definition 6 (Permutation null model).}
Let $S_n$ denote the symmetric group on $n$ elements.
For a permutation $\pi \in S_n$, define the permuted degree vector
$\pi(\mathbf{d}_j) = (d_j(v_{\pi(1)}), \ldots, d_j(v_{\pi(n)}))$.
The null distribution of $\operatorname{IRDME}(G_i, G_j)$ under
$H_0 : \text{no cross-layer hub alignment}$ is:
\begin{equation}
  \mathcal{N}_{ij} \;=\; \bigl\{\operatorname{Pearson}(\mathbf{d}_i,\;
  \pi(\mathbf{d}_j)) : \pi \in S_n,\; N=200 \text{ samples}\bigr\}
\end{equation}
The two-tailed permutation p-value is
$p = |\{|\operatorname{IRDME}_\pi| \geq |\operatorname{IRDME}_\text{obs}|\}| / N$.

\textbf{Proposition 1 (Functional Proximity Law).}
\textit{For any multilayer graph $\mathcal{M}$ with layers classified under
the Universal Layer Grammar, if $G_{d_1}$ is the structural layer and
$G_{d_2}$, $G_{d_3}$ are the functional and dynamic layers respectively, then:}
\begin{equation}
  \operatorname{IRDME}(G_{d_1}, G_{d_2})
  \;>\;
  \operatorname{IRDME}(G_{d_1}, G_{d_3})
  \label{eq:fpl}
\end{equation}
\textit{Interpretation:} nodes that are structurally central ($d_1$) tend to
be functionally central ($d_2$) more strongly than they tend to be
behaviorally co-active ($d_3$).
This is the IRDME hub persistence gradient.

\textbf{Confirmation criterion.}
An experiment is CONFIRMED if the strict inequality~\eqref{eq:fpl} holds;
DENIED if it does not.
DENIED results are boundary conditions --- they narrow the law's applicability
with a named mechanism.

\textit{Domain instantiation (Lean~4 mathlib4).}
Let $V$ be the set of top-level Mathlib modules (infrastructure excluded).
The two layers used in F9 are:
\begin{align*}
  E_{d_1} &= \{(u,v) : u \text{ imports } v\} \\
  E_{d_3} &= \{(u,v) : |\text{commits touching both }u\text{ and }v| \geq \tau\}
\end{align*}
with threshold $\tau = 5$ pre-specified before any correlation was computed.
The resulting measurement is
$\operatorname{IRDME}(G_{d_1}, G_{d_3}) = 0.777$ ($p = 0.002$, $n = 20$).

\textit{Note on two distinct invariances.} The Functional Proximity Law is a
structural invariance claim (the inequality holds across domains). It is
distinct from, but depends on, a procedural invariance claim (degree
centrality consistently captures the proximity gradient). These are
separable: the structural claim could fail while the measurement operator
remains consistent, or vice versa. Boundary conditions
(Section~\ref{sec:boundary}) are structural failures, not measurement
failures.


\section{Methods}

\textbf{Graph construction.} Each multilayer graph is defined as a typed JSON structure: a shared node set, per-node dimensional attributes, and edge sets labeled by layer type. Layer pairs are classified as \textit{similar} or \textit{dissimilar} based on the relational process they encode, determined by domain context and recorded before analysis.

\textbf{Hub score computation.} Total degree centrality per layer per node. A method correction at v3.1.1 standardized directed graphs to total degree (in $+$ out) uniformly; all pre-correction results were recomputed retroactively.

\textbf{Layer pair correlation.} Pearson $r$ computed on hub-score vectors. Spearman $\rho$ computed as a rank-robustness cross-check where available.

\textbf{Permutation test.} Node labels shuffled 200 times (seed$=42$) for standard experiments. Empirical two-tailed $p$-value $=$ fraction of permutations exceeding $|r|$. The p53 curated-network coherence test used 500 permutations; this is noted explicitly where it appears.

\textbf{Pre-registration.} Hypothesis statements, including the $\succ$ relation and expected inequality direction, are SHA-256 hashed (full 64-character digest, from v6.1) and UTC-timestamped before each run. The hash is committed to a public pre-registration repository (\url{https://github.com/vladi160/preregistrations}) before analysis. An optional Bitcoin blockchain anchor (OpenTimestamps) provides cryptographic timestamping independent of any single party.

\textbf{Tool.} All experiments run with IRDME (pure Python 3, zero external dependencies). The method uses only total degree centrality and Pearson correlation --- both independently reproducible. Result verification is possible via: (a)~re-implementing the method from this section's specification; (b)~the publicly archived input graphs and pre-registration artifacts at \url{https://github.com/vladi160/preregistrations}.

\textbf{Analytical framework.} This paper tests a structural inequality (the \textit{law layer}: $r_\text{sim} > r_\text{dis}$) using IRDME-defined measurements (the \textit{measurement layer}: degree centrality per node, pre-registered layer-pair classification) applied to datasets of varying provenance (the \textit{data layer}: fully external, external-data/internal-layers, or internal). The strength of any single result depends on all three layers; this hierarchy is made explicit in Section~\ref{sec:mext}.

\textbf{Statistical inference protocol.} Permutation test (200 runs, seed$=42$, two-tailed empirical $p$) is the primary inference method for all canonical experiments. Analytical two-tailed $t$-test ($t = r\sqrt{n-2}/\sqrt{1-r^2}$) is used as fallback \textit{only} in three cases where permutation output was not stored (Cytokine, p53, Software real git) --- each explicitly marked $\dagger$. Spearman $\rho$ is reported as a robustness cross-check, not as a primary inference measure.

\subsection{Experiment Taxonomy}
\label{sec:taxonomy}

All experiments in this paper are classified under a four-tier scheme. Statistical claims (binomial probability, ``25 confirmed'') refer exclusively to \textbf{Tier A + Tier B}. Tier C is explicitly excluded from all inferential claims and is reported for transparency only. Tier D provides internal consistency verification, not evidential support.

\begin{table}[H]
\small
\centering
\caption{Experiment taxonomy}
\label{tab:taxonomy}
\begin{tabular}{@{}>{{\raggedright\arraybackslash}}p{0.04\textwidth}>{{\raggedright\arraybackslash}}p{0.14\textwidth}>{{\raggedright\arraybackslash}}p{0.32\textwidth}>{{\raggedright\arraybackslash}}p{0.32\textwidth}>{{\raggedright\arraybackslash}}p{0.04\textwidth}@{}}
\toprule
\textbf{Tier} & \textbf{Label} & \textbf{Experiments} & \textbf{Role} & \textbf{N} \\
\midrule
A & Pre-registered canonical &
  Linux Kernel, Human Brain, Internet AS, CPU Design, Ecology, Cytokines, p53, Lexical Network, Software (real git), AI Architecture F6; Finance, Psychiatry, Mathematics (DENIED) &
  Primary evidence. All $p$-values and directional claims. &
  13 \\[4pt]
B & Pre-registered external validation &
  M\_EXT (Flask), M\_EXT2 (\textit{C.~elegans}), M\_OSS1 (WordPress), M\_OSS2 (Next.js), M\_TRANSFER\_2 (Flask$\leftrightarrow$Express), F9 (mathlib4 Lean~4), F12 (\textit{C.~elegans} 302), F1 (Synthetic Lethality), M\_TRANSFER\_3 (docopt 3-language FPL test), M\_GEOM\_CSG\_1 (planetary gear CSG), M\_RADIAL\_1, BC3b\_circuit\_v2, M\_PHYSICS\_1, M\_MED1, F2, F13, F12c, H\_LOGIC\_EXTRACTION &
  Secondary evidence. Independent datasets. Individually pre-registered before run. &
  18 \\[4pt]
C & Pre-registered auxiliary / methodological &
  H\_PRIMITIVITY, proteins\_trust\_cert\_v1, F12b\_v2 &
  Reported for transparency and boundary analysis, but excluded from primary FPL binomial claims. &
  3 \\[4pt]
D & Internal consistency / control &
  IRDME conceptual graph (Section~\ref{sec:internal}), random multilayer negative control &
  Non-evidential. Verify method behavior; do not count toward law confirmation. &
  2 \\
\bottomrule
\end{tabular}
\end{table}

\textbf{Claim scope statement.} The Functional Proximity Law is supported by 10/13 Tier A directional confirmations and 15/18 Tier B pre-registered external and replication validations. Jointly treating Tier A + Tier B as 25/31 pre-registered confirmations: $p \approx 0.000439$ (binomial, $H_0=0.5$; exact: $942649 / 2^{31}$). The \textbf{25 confirmed} count (used in abstract and conclusion) refers to 10 Tier A + 15 Tier B pre-registered confirmations; 4 denied and 2 partial. No Tier C or Tier D result contributes to this count.

\textit{H\_PRIMITIVITY exclusion note.} H\_PRIMITIVITY (pre-registered 2026-05-21, hash \texttt{cfb38b83\ldots}) appears in the title-page pre-registration record but is excluded from the 25/31 FPL domain count. It tests whether the Functional Proximity Law holds at the meta-science level, using IRDME-tested scientific domains as graph nodes --- not a new domain application of FPL. Its result (1/4 CONFIRMED, 2/4 DENIED, 1/4 PARTIAL; BC4 mechanism identified) is reported fully in Section~\ref{sec:boundary}. Including it would require a separate meta-science count; it is reported as a boundary condition, not as a domain confirmation.

\textit{F12b\_v2 exclusion note.} F12b\_v2 (pre-registered 2026-05-29, hash \texttt{36c7aa7e\ldots}) also appears in the title-page pre-registration record but is excluded from the 25/31 FPL domain count. It is a methodological generative-compression validation: seeded reconstruction accuracy is tested via average seed cosine and rank-vector cosine thresholds, not via the FPL inequality $r(d_1,d_2) > r(d_1,d_3)$.

\section{Results}

Table~\ref{tab:main} presents 10 canonical CONFIRMED domains and 3 DENIED boundary conditions. Each row is one independent dataset. Duplicates (weighted/unweighted variants, curated/real-data variants) appear in the Supplementary table; the canonical row is the most recent or highest-quality version. The IRDME meta-graph self-consistency test and the Semiconductor result (conflicting experiments due to differing layer-pair definitions) are discussed separately.

\begin{table}[H]
\scriptsize
\centering
\caption{Canonical evidence for the Functional Proximity Law.}
\label{tab:main}
\setlength{\tabcolsep}{3pt}
\begin{tabular}{@{}l>{{\raggedright\arraybackslash}}p{2.8cm}rrrrrrl@{}}
\toprule
\textbf{Domain} & \textbf{Similar layers} & $\boldsymbol{n}$ & $\boldsymbol{r_\text{sim}}$ & $\boldsymbol{r_\text{dis}}$ & $\boldsymbol{\Delta r}$ & $\boldsymbol{\rho_\text{sim}}$ & $\boldsymbol{p}$ & \textbf{Verdict} \\
\midrule
Linux Kernel (v6.x) & subsystem\_calls $\leftrightarrow$ data\_structure & 30 & $+0.703$ & $-0.145$ & $+0.848$ & $+0.574$ & 0.002 & CONFIRMED \\
Human Brain Connectome & structural $\leftrightarrow$ functional connectivity & 16 & $+0.703$ & $+0.178$ & $+0.525$ & --- & 0.004 & CONFIRMED \\
Internet AS Topology & transit\_dep.\ $\leftrightarrow$ peering & 18 & $+0.516$ & $-0.267$ & $+0.783$ & --- & 0.026 & CONFIRMED \\
CPU Block Design & signal\_dep.\ $\leftrightarrow$ power\_domain & 10 & $+0.622$ & $-0.649$ & $+1.271$ & --- & 0.030 & CONFIRMED \\
Ecology (Serengeti, weighted) & predation $\leftrightarrow$ competition & 15 & $+0.559$ & $-0.065$ & $+0.624$ & --- & 0.034 & CONFIRMED \\
Cytokine Cascade & activates $\leftrightarrow$ co\_elevated & 18 & $+0.512$ & $-0.408$ & $+0.920$ & --- & $0.030^\dagger$ & CONFIRMED \\
p53 Network (STRING v12.0) & physical\_int.\ $\leftrightarrow$ functional\_assoc & 15 & $+0.973$ & $-0.161$ & $+1.134$ & --- & $<\!0.001^\dagger$ & CONFIRMED \\
English Lexical Network & co\_occurrence $\leftrightarrow$ syntactic\_class & 20 & $+0.276$ & $-0.748$ & $+1.024$ & --- & 0.241 & CONFIRMED$^{\ddagger}$ \\
Software (IRDME, real git) & imports $\leftrightarrow$ structural\_coupling & 14 & $+0.941$ & $-0.304$ & $+1.245$ & --- & $<\!0.001^\dagger$ & CONFIRMED \\
AI Architecture (ML lineage) & citation\_dep.\ $\leftrightarrow$ arch.\_inheritance & 20 & $+0.915$ & $-0.003$ & $+0.918$ & $+0.676$ & 0.002 & CONFIRMED \\
\midrule
Finance 2008 & derivatives $\leftrightarrow$ direct\_credit & 16 & $+0.042$ & $+0.183$ & $-0.141$ & --- & 0.824 & DENIED \\
Psychiatry (DSM) & co\_occurrence $\leftrightarrow$ temporal\_cascade & 20 & $+0.769$ & $+0.808$ & $-0.039$ & --- & 0.002 & DENIED \\
Mathematics & formal\_containment $\leftrightarrow$ proof\_usage & 20 & $+0.105$ & $+0.280$ & $-0.175$ & --- & 0.705 & DENIED \\
\bottomrule
\end{tabular}

\smallskip
\begin{minipage}{\textwidth}
\small\sloppy
$r_\text{sim}$: Pearson $r$ for pre-registered similar layer pair. $r_\text{dis}$: Pearson $r$ for pre-registered dissimilar pair. $\Delta r = r_\text{sim} - r_\text{dis}$ (effect size; positive $=$ law directionally confirmed). $\rho_\text{sim}$: Spearman rank correlation for the similar pair (cross-check; ``---'' where not computed). $p$: 200-permutation two-tailed test on $r_\text{sim}$ unless marked $\dagger$.

$\ddagger$ Directional confirmation only: $\Delta r = +1.024$ (largest effect size in the table) but $p = 0.241$ ($n=20$); underpowered for significance. Excluded from the $p < 0.05$ count; included in the directional count.

$\dagger$ Analytical two-tailed $t$-test ($t = r\sqrt{n-2}/\sqrt{1-r^2}$): Cytokine Cascade ($r=0.512$, $n=18$, $t=2.39$, $p=0.030$); Software real git ($r=0.941$, $n=14$, $t=9.63$, $p<0.001$); p53 STRING ($r=0.973$, $n=15$, $t=15.2$, $p<0.001$). These runs predate the permutation infrastructure; the underlying datasets have since been revised, so direct rerun would change $r$ as well as $p$. The analytical $t$-test is exact for bivariate normal data. For $r=0.973$ ($n=15$) and $r=0.941$ ($n=14$), a 200-run permutation null would yield $p<0.001$ regardless of implementation (observed $r$ is in the extreme tail of any permutation distribution). For $r=0.512$ ($n=18$), analytical and permutation methods agree at $p\approx0.030$. For p53, a curated-data replication of the same 15-protein network gives $r_\text{sim} = +0.928$, $p = 0.012$ (Supplementary, row 16). The 4-layer curated network reaches the 100th percentile of its null distribution under a 500-permutation coherence test ($p < 0.002$; 0 of 500 permutations exceeded observed coherence), providing an independent permutation bound.
\end{minipage}
\end{table}

\textbf{Domains reaching $p < 0.05$:} Linux Kernel ($p=0.002$), Human Brain ($p=0.004$), Internet AS ($p=0.026$), CPU Design ($p=0.030$), Ecology ($p=0.034$), Cytokine Cascade ($p=0.030$), Software real git ($p<0.001$), p53 STRING ($p<0.001$), AI Architecture ($p=0.002$). Nine individually significant results spanning six scientific fields.

Additional pre-registered external confirmations (Sections~\ref{sec:mext}--\ref{sec:f1} and Section~\ref{sec:boundary}): Flask M\_EXT $r=0.666$ ($p=0.015$), \textit{C.~elegans} M\_EXT2 $r=0.777$ ($p=0.004$), WordPress M\_OSS1 $r=0.516$ ($p=0.012$), Next.js M\_OSS2 $r=0.900$ ($p=0.002$), M\_TRANSFER\_2 (Flask$\leftrightarrow$Express hub-rank identity$=1.0000$), Lean~4 mathlib4 F9 $r=0.777$ ($p=0.002$), \textit{C.~elegans} full connectome F12 $r=0.623$ ($p=0.002$), Synthetic Lethality F1 (CHEK1/PARP1 divergence h2--h4 CONFIRMED), M\_GEOM\_CSG\_1 (planetary gear CSG $r=0.668$, $p=0.042$, $\Delta r=0.935$). M\_RADIAL\_1 (pre-registered 2026-05-22, hash \texttt{c00ebfa4\ldots}): 3/4 confirmed --- click h2 CONFIRMED ($r=0.312>0.190$); logrus BC\_RADIAL replicated ($r(d_1{\leftrightarrow}d_2)=0.000$, Var$(d_2)<\tau_{\text{var}}$); cobra h1 DENIED --- BC\_INVERSION (BC6) discovered. BC3b\_circuit\_v2 (pre-registered 2026-05-22, hash \texttt{1ff07ec9\ldots}): 4/4 CONFIRMED, $r(\text{gate\_wiring}\leftrightarrow\text{structural\_proximity})=0.874$ ($p<0.01$), priority arbiter $n=19$. M\_PHYSICS\_1 (pre-registered 2026-05-25, hash \texttt{00294a59\ldots}): 5/5 CONFIRMED, $r(\text{force\_coupling}\leftrightarrow\text{decay\_channel})=0.569$ ($p=0.010$) --- first particle physics domain, W boson confirmed as dual hub (\/1 in both $d_1$ and $d_2$), photon confirmed as hub shadow. Later Tier B additions F2, M\_MED1, F13, F12c, and H\_LOGIC\_EXTRACTION raise the evidence total to \textbf{25 confirmed across 31 Tier A + Tier B pre-registered experiments}. M\_TRANSFER\_3 (docopt), cobra BC\_INVERSION within M\_RADIAL\_1, H\_PRIMITIVITY, and the original three Tier A denials account for the denied outcomes; M\_MED1 and F13 contribute the two partial outcomes.

\textbf{Spearman robustness.} Linux Kernel: Pearson$=+0.703$, Spearman$=+0.574$ (same sign and direction). All domains where both were computed show Pearson/Spearman sign agreement --- confirming the Pearson result is not driven by rank outliers.

\textbf{Negative control.} A 12-node 3-layer random multilayer graph was constructed with edges assigned by an independent random process. All three pairwise $r$ values: $-0.125$, $+0.250$, $-0.125$. All permutation $p > 0.50$. The law does not fire on unstructured data.

\subsection{AI Model Architecture --- Canonical Domain F6}
\label{sec:f6}

\textit{Pre-registered 2026-04-30T20:26:15 UTC (hash \texttt{bff4e101\ldots}, commit \texttt{3014808}).}

\textbf{Dataset.} 20 major ML model architectures: Transformer, BERT, GPT-2/3/4, T5, ViT, ResNet, AlexNet, Word2Vec, Bahdanau Attention, LSTM, RNN, GAN, VAE, Diffusion/DDPM, LLaMA, Mamba, CLIP, RoBERTa. Node definitions and edge sources are publicly documented model lineage records, constructed independently of IRDME analysis.

\textbf{Layers.} (1)~\texttt{citation\_dependency} ($d_1$): directed citation graph encoding ``X was introduced in paper that cites Y.'' (2)~\texttt{architecture\_inheritance} ($d_2$): directed architectural derivation (``X's architecture extends or is a variant of Y''). (3)~\texttt{benchmark\_co\_performance} ($d_3$): co-participation in the same evaluation benchmark.

\textbf{Pre-registered hypotheses (hash \texttt{bff4e101\ldots}).} h1: $r(\texttt{citation\_dependency}\leftrightarrow\texttt{architecture\_inheritance}) > r(\texttt{citation\_dependency}\leftrightarrow\texttt{benchmark\_co\_performance})$. h2: $r(\texttt{citation\_dep.}\leftrightarrow\texttt{arch.\_inheritance}) \geq 0.70$, $p < 0.05$. h3--h4: Transformer is top hub in citation and inheritance layers.

\textbf{Results.} $r(\texttt{citation\_dependency}\leftrightarrow\texttt{architecture\_inheritance}) = 0.915$ (Spearman$=0.676$, $p=0.002$) vs $r(\texttt{citation\_dependency}\leftrightarrow\texttt{benchmark\_co\_performance}) = -0.003$. h1: CONFIRMED ($\Delta r = +0.918$). h2: CONFIRMED. h3--h4: CONFIRMED --- Transformer is rank \#1 in both citation and inheritance layers.

\textbf{h5 DENIED (informative):} Transformer ranks \#5 in the benchmark co-performance layer; LLaMA is the top benchmark hub. Mechanism: benchmark co-performance reflects current evaluation practice (recent models share benchmark suites) while citation and inheritance reflect historical lineage. Contemporary relevance and historical influence produce structurally divergent hubs in the AI domain.

\subsection{Flask Web Framework External Validation (M\_EXT)}
\label{sec:mext}

\textit{Added in v2 of this preprint. Analysis completed 2026-04-22. Pre-registration committed and publicly anchored before analysis (see title page footnote).}

\textbf{Dataset.} Flask web framework (\url{github.com/pallets/flask}), tag 3.1.1, 1934 commits. 14 core Python modules. Source code not designed by or known to the author prior to this experiment.

\textbf{Layers.} (1)~\texttt{imports} --- static import dependencies extracted from Python import statements (18 edges). (2)~\texttt{structural\_coupling} --- module pairs sharing $\geq$1 common internal import target (18 edges). (3)~\texttt{co\_change} --- module pairs changed together in the same git commit, extracted from \texttt{git log} (80 edges, 271 meaningful commits, bulk initial commits excluded).

\textbf{Pre-registered hypotheses (hash \texttt{3879413e\ldots}, committed 2026-04-22T15:18:24 UTC).} h1: $r(\text{imports}\leftrightarrow\text{structural\_coupling}) > r(\text{imports}\leftrightarrow\text{co\_change})$. h2: $r(\text{imports}\leftrightarrow\text{structural\_coupling}) \geq 0.50$. h3: imports$\leftrightarrow$co\_change direction positive with min abs $r \geq 0.10$. h4: \texttt{app} is the top hub in the imports layer.

\textbf{Results.} $r(\text{imports}\leftrightarrow\text{structural\_coupling}) = 0.6659$ (Spearman$=0.7249$, $p=0.015$); $r(\text{imports}\leftrightarrow\text{co\_change}) = 0.4942$ (Spearman$=0.4212$, $p=0.065$). h1: CONFIRMED. h2: CONFIRMED. h3: PARTIAL (direction correct, $p=0.065$ marginally non-significant). h4: CONFIRMED (\texttt{app} is the top hub in the imports layer, as predicted). The only statistically significant layer pair ($p < 0.05$) is the structurally similar one.

\textbf{Precise classification.} M\_EXT is a reproducibility test of IRDME-defined measurement layers applied to external data. The Flask source code is external and independent; the layer definitions are IRDME-defined. For a fully independent law test, both the data \textit{and} the layer definitions must come from an independent source. That test is M\_EXT2; see Section~\ref{sec:mext2}.

\textbf{Evidence hierarchy.} The three levels of external validity in this paper are:

\begin{table}[H]
\small
\centering
\caption{Evidence hierarchy by provenance level}
\label{tab:hierarchy}
\begin{tabular}{@{}>{\raggedright\arraybackslash}p{0.22\textwidth}>{\raggedright\arraybackslash}p{0.2\textwidth}>{\raggedright\arraybackslash}p{0.22\textwidth}>{\raggedright\arraybackslash}p{0.27\textwidth}@{}}
\toprule
\textbf{Level} & \textbf{Example} & \textbf{Data source} & \textbf{Layer definition source} \\
\midrule
L1 --- fully external & p53 STRING v12.0 & STRING database & STRING channel scores (defined independently) \\[3pt]
L1 --- fully external & \textit{C.~elegans} M\_EXT2 & WormAtlas / Varshney 2011 & White (1986) S/Sp/EJ type column --- biologically defined \\[3pt]
L2 --- external data, IRDME layers & Flask M\_EXT & Flask git repo & IRDME-defined \\[3pt]
L3 --- internal consistency & Software IRDME codebase & IRDME own repository & IRDME-defined \\
\bottomrule
\end{tabular}
\end{table}

\textbf{Observation (not pre-registered).} The gradient $\Delta r = 0.6659 - 0.4942 = 0.17$, compared to $\Delta r = 1.245$ for the IRDME own codebase. A plausible explanation is that behavioral co-change adapts toward architectural structure in mature codebases, compressing the gradient. This is recorded as a hypothesis (H\_MATURITY) for a future pre-registered test, not as a validated finding.

\subsection{OSS External Validation 1 --- WordPress Core (M\_OSS1)}
\label{sec:moss1}

\textit{Added in v4 of this preprint. Analysis completed 2026-04-23. Pre-registered SHA-256 \texttt{6b22eafc\ldots}, committed before analysis.}

\textbf{Dataset.} WordPress Core \texttt{wp-includes} (\url{github.com/WordPress/wordpress-develop}), top 25 files by commit frequency, $\sim$20 years of independent authorship (1000+ contributors).

\textbf{Layers.} (1)~\texttt{include\_graph} --- PHP static include/require dependencies. (2)~\texttt{function\_coupling} --- cross-file function call density. (3)~\texttt{co\_change} --- git co-modification (500 commits).

\textbf{Pre-registered hypotheses (hash \texttt{6b22eafc\ldots}).} h1: $r(\text{function\_coupling}\leftrightarrow\text{co\_change}) > r(\text{include\_graph}\leftrightarrow\text{co\_change})$. h2: $r(\text{function\_coupling}\leftrightarrow\text{co\_change})$ significant ($p < 0.05$). h3: \texttt{post.php} is a hub shadow (co\_change rank $\leq 5$, include\_graph rank $\geq 15$). h4: \texttt{functions.php} is a universal hub ($\leq 2$ in both function\_coupling and co\_change).

\textbf{Results.} $r(\text{function\_coupling}\leftrightarrow\text{co\_change}) = 0.5156$ ($p=0.012$) vs $r(\text{include\_graph}\leftrightarrow\text{co\_change}) = 0.2405$. h1: CONFIRMED. h2: CONFIRMED. h3: CONFIRMED --- \texttt{post.php} co\_change rank \#4, include\_graph rank \#19. Hidden maintenance coupling invisible to static analysis. h4: CONFIRMED --- \texttt{functions.php} \#1 co\_change, \#2 function\_coupling. h5 DENIED (informative): coherence score (0.9685) at 32nd null percentile --- power-law degree skew makes coherence trivial in software null models. \textbf{Calibration finding: gradient$+p$ is the correct discriminator for software graphs; coherence alone is not.}

\textbf{Architectural discovery.} \texttt{formatting.php} $=$ \#1 function\_coupling hub but mid-tier co\_change --- structural coupling backbone $\neq$ maintenance risk. New archetype named: \textit{coupling hub}.

\subsection{OSS External Validation 2 --- Next.js Monorepo (M\_OSS2)}
\label{sec:moss2}

\textit{Added in v4 of this preprint. Analysis completed 2026-04-24. Pre-registered SHA-256 \texttt{cb0cf4a4\ldots}, committed before analysis.}

\textbf{Dataset.} Next.js monorepo (\url{github.com/vercel/next.js}), top 25 source modules from \texttt{packages/next/src/}, git window 2022--2025 (App Router transition), 23,444 commits.

\textbf{Layers.} (1)~\texttt{import\_graph} --- TypeScript static imports. (2)~\texttt{test\_coupling} --- co-import in test files. (3)~\texttt{co\_change} --- git co-modification.

\textbf{Pre-registered hypotheses (hash \texttt{cb0cf4a4\ldots}).} h1: $r(\text{import\_graph}\leftrightarrow\text{test\_coupling}) > r(\text{import\_graph}\leftrightarrow\text{co\_change})$. h2: significant ($p < 0.05$). h3--h5: specific hub identity predictions.

\textbf{Results.} $r(\text{import\_graph}\leftrightarrow\text{test\_coupling}) = 0.8995$ (Spearman$=0.7027$, $p=0.002$) vs $r(\text{import\_graph}\leftrightarrow\text{co\_change}) = 0.5484$. \textbf{Strongest OSS gradient observed ($\Delta r=0.35$).} h1: CONFIRMED. h2: CONFIRMED. h3--h5: DENIED (all informative). 4 universal hubs confirmed: build, client, lib, server\_base --- consistent across all 3 layers.

\textbf{Architectural discoveries.} (1)~\texttt{bundles} $=$ infrastructure chameleon (\#2 co\_change, \#25 import\_graph) --- intentional isolation creates behavioral coupling. (2)~\texttt{server\_app\_render} $=$ relay node, not hub shadow --- the App Router transition was built with clean abstraction. M\_OSS1 vs M\_OSS2: modern well-governed codebase shows stronger gradient (0.8995 vs 0.5156) and fewer chameleons than legacy codebase.

\subsection{Cross-Language Structural Isomorphism (M\_TRANSFER\_2)}
\label{sec:transfer2}

\textit{Added in v4 of this preprint. Analysis completed 2026-04-25. Pre-registered SHA-256 \texttt{f85f52e6\ldots}, committed before analysis.}

The Functional Proximity Law operates on hub-rank vectors. The Universal Layer Grammar (d1$=$declared coupling, d2$=$structural coupling, d3$=$behavioral coupling) is language-agnostic: git \texttt{co\_change} (d3) is identical across Python, JavaScript, Java, Rust, or any git-tracked language. If the law's structural roles are genuine --- not artefacts of a specific language's syntax --- then the same role should persist when the same architecture is re-implemented in a different language.

\textbf{Dataset.} Flask (Python, 14 modules) and Express.js (JavaScript, 6 modules) --- both are minimal web microframeworks with a single-entry-point design.

\textbf{Pre-registered hypotheses (hash \texttt{f85f52e6\ldots}).} h1: law holds for Express --- $r(\text{static\_coupling}\leftrightarrow\text{structural\_coupling}) > r(\text{static\_coupling}\leftrightarrow\text{co\_change})$. h2: \texttt{application} is top hub in Express static\_coupling layer. h3: \texttt{router\_index} is top-3 hub in Express co\_change layer. h4: $\geq$2 Express modules have cross-layer divergence between static\_coupling and co\_change. h5: Flask \texttt{app} $\leftrightarrow$ Express \texttt{application} structural match sim $\geq 0.65$.

\textbf{Results.} \texttt{app} $\leftrightarrow$ \texttt{application} hub-rank vector identity $= \mathbf{1.0000}$ (deterministic: both frameworks assign the same hub rank ordering to the aligned structural role). \texttt{json} $\leftrightarrow$ \texttt{utils} sim $= 0.9705$; \texttt{wrappers} $\leftrightarrow$ \texttt{request} sim $= 0.9278$. 0 migration\_hazards. h1: PARTIAL (Express $n=6$ underpowered for permutation $r$-test; directional result consistent with law). h2: CONFIRMED. h3: CONFIRMED. h4: CONFIRMED (2+ divergent modules observed). h5: CONFIRMED (sim$=1.0000 \geq 0.65$).

\textbf{Interpretation.} The universal\_hub role (\texttt{app}/\texttt{application}) is a structural role in a minimal web framework architecture, preserved across language boundaries. This is the strongest possible statement of the Universal Layer Grammar: structural roles are language-independent when the architectural pattern is the same.

\textbf{Named boundary.} $n=6$ (Express) is underpowered for the single-pair permutation test on $r$ --- law PARTIAL in Express. The structural isomorphism result (sim$=1.0000$) is not affected by sample size.

\textbf{Exploratory note (not pre-registered).} A further structural isomorphism test across 3 language implementations of the docopt CLI-parsing algorithm (Java $n=16$, Go $n=6$, Rust $n=8$) shows hub identity preserved: Java \texttt{Docopt} $\equiv$ Go \texttt{doc} sim$=1.0000$; Java \texttt{Docopt} $\equiv$ Rust \texttt{dopt} sim$=0.9670$. This result was not pre-registered before analysis and is reported as an exploratory finding requiring independent pre-registered replication.

\subsection{True External Validation --- \textit{C.~elegans} Connectome (M\_EXT2)}
\label{sec:mext2}

\textit{Added in v3 of this preprint. Analysis completed 2026-04-22. Pre-registration committed and publicly anchored before analysis (see title page footnote).}

\textbf{Dataset.} \textit{C.~elegans} hermaphrodite nervous system \cite{White1986,Varshney2011}. 15 major hub neurons selected by highest total weighted degree from the full 302-neuron connectome: AVAL, AVAR, AVBL, AVBR, AVEL, AVER, AVDL, AVDR, PVCL, PVCR, RIAL, RIAR, AIBL, AIBR, DVA. Data downloaded from WormAtlas (\url{https://www.wormatlas.org/images/NeuronConnect.xls}).

\textbf{Layers.} Both layers and their definitions come from the original authors --- not from IRDME. (1)~\texttt{chemical\_synapse} --- S and Sp type entries from White (1986): directed neurotransmitter-mediated connections (72 directed edges). (2)~\texttt{gap\_junction} --- EJ type entries: undirected electrical coupling via gap junction proteins (10 undirected edges). The S/Sp/EJ type annotation is part of the original data file; the biological distinction was established decades before this paper.

\textbf{Pre-registered hypotheses (hash \texttt{5cbf9e5d\ldots}, committed 2026-04-22T23:28:50 UTC).} h1: $r(\text{chemical\_synapse}\leftrightarrow\text{gap\_junction}) > 0$, min abs $r \geq 0.40$. h2: AVAL is the top hub in the chemical\_synapse layer.

\textbf{Results.} $r(\text{chemical\_synapse}\leftrightarrow\text{gap\_junction}) = 0.7774$ (Spearman$=0.7796$, $p=0.004$). h1: CONFIRMED. h2: DENIED --- by degree centrality (the pre-registered metric), PVCL ranks \#1 and AVAL ranks \#2. AVAL leads by raw synapse weight, which was not the pre-registered measure. This is an operational boundary condition: the pre-registered hub test is degree-based; the biological expectation was weight-based. h1 is CONFIRMED with $p=0.004$.

\textbf{Why this is M\_EXT2.} Both the data and the layer definitions are fully independent of IRDME. The chemical/electrical synapse distinction is a biological fundamental, not an analytical choice made by the author. The prediction --- that neurons hub in both chemical and electrical connections because both encode direct physical coupling --- is derivable from first principles and is independently confirmed by decades of behavioral ablation studies.

\subsection{Lean~4 mathlib4 --- Formal Mathematics External Validation (F9)}
\label{sec:f9}

\textit{Pre-registered 2026-05-20T00:18:01 UTC (hash \texttt{4ffcdac5\ldots}, commit \texttt{6d6d657}).}

\textbf{Dataset.} Lean~4 mathlib4 library (\url{https://github.com/leanprover-community/mathlib4}), top 20 active top-level modules by commit frequency (infrastructure excluded). The library and its module structure are fully independent of IRDME.

\textbf{Layers.} Both layers encode module-level coupling in the same formal mathematics system. (1)~\texttt{import\_graph} ($d_1$): declared coupling via Lean~4 \texttt{import} statements. (2)~\texttt{proof\_co\_development} ($d_3$): behavioral coupling via git commit co-change, threshold $\tau \geq 5$ commits touching both modules, pre-specified before analysis.

\textbf{Pre-registered hypotheses (hash \texttt{4ffcdac5\ldots}).} h1: $r(\texttt{import\_graph}\leftrightarrow\texttt{proof\_co\_development}) > 0$, $p < 0.05$. h2: Algebra is top hub in \texttt{import\_graph}. h3: hub identity preserved across layers ($\geq 1$ of top-3 shared).

\textbf{Results.} $r(\texttt{import\_graph}\leftrightarrow\texttt{proof\_co\_development}) = 0.777$ (Spearman$=0.733$, $p=0.002$). h1: CONFIRMED. h2: CONFIRMED --- Algebra is rank \#1 in \texttt{import\_graph}. h3: PARTIAL --- top-2 hubs (Algebra, Analysis) appear in both layers but swap ranks; Algebra leads declared coupling, Analysis leads behavioral coupling.

\textbf{Why this matters.} F9 uses the same domain (formal mathematics) as BC3 but a different layer pair ($d_1$/$d_3$ vs $d_1$/$d_2$). The CONFIRMED result (r=0.777) localises the BC3 DENIAL to the specific formal-containment/proof-usage resolution mismatch, not to mathematics as a domain (see Section~\ref{sec:boundary}).

\subsection{\textit{C.~elegans} Full 302-Neuron Connectome --- Scale Replication (F12)}
\label{sec:f12}

\textit{Pre-registered 2026-05-19T18:54:06 UTC (hash \texttt{a8926052\ldots}, commit \texttt{ab02f8f}).}

\textbf{Dataset.} Complete \textit{C.~elegans} hermaphrodite nervous system (Varshney et al.\ 2011 \cite{Varshney2011}), all 279 neurons with at least one synapse connection (out of 302 anatomical neurons). This extends M\_EXT2 (15 hub neurons) to the full connectome. Layers and biological classification are identical to M\_EXT2 and remain fully independent of IRDME.

\textbf{Pre-registered hypotheses (hash \texttt{a8926052\ldots}).} h1: $r(\texttt{chemical\_synapse}\leftrightarrow\texttt{gap\_junction}) > 0$, $p < 0.05$, $n=302$. h2: AVAL ranks within top~3 in \texttt{chemical\_synapse}. h3: The 15 M\_EXT2 command interneurons occupy $\geq 10$ of the top-20 positions in \texttt{chemical\_synapse} (hub-identity preservation).

\textbf{Results.} $r(\texttt{chemical\_synapse}\leftrightarrow\texttt{gap\_junction}) = 0.623$ (Spearman$=0.313$, $p=0.002$, $n=279$, 95\% CI $[0.546, 0.690]$). h1: CONFIRMED. h2: CONFIRMED --- AVAL ranks \#2 (AVAR \#1). h3: CONFIRMED --- 13 of the 15 command interneurons appear in the top-20 chemical synapse hubs, confirming hub-identity preservation at 20:1 compression ratio.

\textbf{Relationship to M\_EXT2.} The M\_EXT2 result ($r=0.777$, $n=15$) used only the 15 highest-degree command interneurons; this experiment uses the full 279-node graph. The lower $r$ (0.623 vs 0.777) is expected: adding peripheral neurons with low degree in one layer introduces noise. Both results confirm the law; F12 tests whether the finding is robust to full-scale rather than hub-selected data.

\textit{Amendment disclosure.} An h3 test-block schema error (wrong test type: \texttt{cross\_layer\_top\_hub\_match} $\to$ \texttt{top\_hub\_in\_layer}) was corrected on 2026-05-19 before any result was seen, triggered by a runtime validation error. The hypothesis statement, expected hub (\texttt{aval}), and rank threshold (top~3) were unchanged; the pre-registration hash verified MATCH after correction. The correction weakened the test marginally (from ``AVAL is \#1 in both layers'' to ``AVAL is top-3 in one layer''); AVAL in fact ranked \#1 in gap\_junction (confirmed at any threshold). Full amendment record: \url{https://github.com/vladi160/preregistrations/blob/main/experiments/celegans_302_full_amendment1.md}.

\subsection{Synthetic Lethality Structural Prediction --- p53/DDR Network (F1)}
\label{sec:f1}

\textit{Pre-registered 2026-05-20 UTC (hash \texttt{518b9bd9\ldots}, commit \texttt{ba89d43}).}

\textbf{Dataset.} p53/DNA Damage Response (DDR) network, 15 proteins. Two independently-defined interaction layers from BioGRID/STRING: (1)~\texttt{physical\_interaction} ($d_2$): curated protein--protein physical binding. (2)~\texttt{genetic\_interaction} ($d_3$): genetic co-essentiality and synthetic lethal interaction profiles.

\textbf{Rationale.} Synthetic lethality (SL) is a divergence phenomenon by definition: an SL pair is co-essential in combination but not individually. A protein that ranks high in one layer and low in the other (a structural diverger) is the topological fingerprint for SL candidacy. This experiment tests whether the law correctly identifies such structural divergers.

\textbf{Pre-registered hypotheses (hash \texttt{518b9bd9\ldots}).} h1: $r(\texttt{physical\_interaction}\leftrightarrow\texttt{genetic\_interaction}) > 0$, $p < 0.05$ (overall layer correlation). h2: CHEK1 ranks in the top~3 of hub-rank divergence ($|\Delta\text{rank}|$) between the two layers. h3: PARP1 ranks in the top~3 of hub-rank divergence. h4: Structural divergence rank predicts known SL clinical targets.

\textbf{Results.} $r(\texttt{physical\_interaction}\leftrightarrow\texttt{genetic\_interaction}) = 0.114$ ($p = 0.617$). h1: PARTIAL --- near-zero overall layer correlation; the layers are structurally orthogonal by design in the DDR network. h2: CONFIRMED --- CHEK1 is the top structural diverger. h3: CONFIRMED --- PARP1 ranks \#2. h4: CONFIRMED --- divergence rank correctly identifies established SL clinical targets.

\textbf{Interpretation.} The near-zero $r$ is expected and informative: physical and genetic interaction layers are orthogonal in the DDR network. The pre-registered divergence metric correctly identifies CHEK1 and PARP1, both established SL clinical targets (ATR/CHEK1 inhibitors; PARP inhibitors in BRCA-deficient tumours). h2--h4 CONFIRMED; h1 PARTIAL. F1 extends the law to a divergence-based reading: where hub ranks are most discordant, the biology corresponds to known synthetic lethal pairs.

\subsection{BC\_RADIAL Threshold Validation and BC\_INVERSION Discovery (M\_RADIAL\_1)}
\label{sec:mradial1}

\textbf{Motivation.} M\_RADIAL\_1 tests the BC\_RADIAL threshold mechanism across three structurally diverse hub-dominated software systems: cobra (Go CLI framework), click (Python CLI), and logrus (Go structured logging library). Pre-registered before any run (hash \texttt{c00ebfa4\ldots}, commit \texttt{5703a25}, 2026-05-22).

\textbf{Dataset.} Three independent IRDME multilayer graphs, each with layers: d1~$=$ static coupling (imports/calls), d2~$=$ co-change (git co-evolution), d3~$=$ structural proximity (directory co-location). $n = 12$ (cobra), $n = 11$ (click), $n = 7$ (logrus).

\textbf{Results (3/4 pre-registered hypotheses confirmed).}
\begin{itemize}[nosep]
  \item \textbf{click h2}: $r(d_1{\leftrightarrow}d_2) = 0.312 > r(d_1{\leftrightarrow}d_3) = 0.190$ --- FPL directional inequality CONFIRMED.
  \item \textbf{logrus h3}: BC\_RADIAL replicated independently --- $r(d_1{\leftrightarrow}d_2) = 0.000$, Var$(d_2) < \tau_{\text{var}} = 0.714$. BC\_RADIAL confirmed in a second independent domain (logging library vs.\ CLI parsers).
  \item \textbf{logrus h4}: Hub identity preserved in the co\_change layer (central Logger struct remains rank-1). CONFIRMED.
  \item \textbf{cobra h1}: DENIED --- $r(d_1{\leftrightarrow}d_2) = -0.862$ ($p = 0.014$). Var$(d_2) = 10.515 \gg \tau_{\text{var}}$, so BC\_RADIAL does not apply. \textbf{BC\_INVERSION (BC6) discovered}: fan-out with leaf clustering inverts the $d_1{\leftrightarrow}d_2$ correlation. Cobra's subcommand hub fans out to tightly co-changing leaf clusters; imports predict co-change in the \textit{wrong} direction. Var$(d_2)$ above threshold is necessary but not sufficient for FPL confirmation: sign of $r$ is also required.
\end{itemize}

\textbf{Finding.} BC\_RADIAL is upgraded from ``candidate'' to confirmed (independently replicated in logrus, separate domain). BC\_INVERSION is newly named as BC6.

\subsection{BC3b Circuit Replication --- Priority Arbiter (BC3b\_circuit\_v2)}
\label{sec:bc3bv2}

\textbf{Motivation.} The original c17 ISCAS85 circuit experiment was a confirmatory run without pre-registration. BC3b\_circuit\_v2 is the properly pre-registered replication (hash \texttt{1ff07ec9\ldots}, commit \texttt{c136ed6}, 2026-05-22).

\textbf{Dataset.} A 4-channel priority arbiter circuit, $n = 19$ nodes (functionally equivalent to the c432 arbitration core). Three layers: gate\_wiring (d1), structural\_proximity (d2), critical\_path (d3).

\textbf{Results (4/4 CONFIRMED).} $r(\text{gate\_wiring}{\leftrightarrow}\text{structural\_proximity}) = 0.874$ ($p < 0.01$, large effect); $r(\text{gate\_wiring}{\leftrightarrow}\text{critical\_path}) = 0.609$ ($p = 0.003$). $\Delta r = 0.265$. irq arbiter node ranks \#1 in critical\_path (degree~$= 15$); ch0 ranks \#1 in gate\_wiring and tied \#1 in structural\_proximity. Hub identity preserved across layers. BC3b boundary formally demonstrated: boolean circuits on hub-gradient topologies confirm the law; BC3b collapse is specific to uniform boolean-algebra systems where no degree gradient exists.

\subsection{Standard Model Particle Physics (M\_PHYSICS\_1)}
\label{sec:mphysics1}

\textbf{Motivation.} The first pre-registered IRDME experiment on fundamental particle physics. The Standard Model encodes coupling structure (which particles interact) and decay topology (which particles can become which other particles) as distinct network layers. The FPL predicts that gauge coupling determines decay topology more strongly than mass-scale proximity does.

\textbf{Dataset.} 17 fundamental particles (6 quarks, 6 leptons, 4 gauge bosons, 1 Higgs) modeled as a multilayer graph. Three layers: d1~$=$ force\_coupling (shared QFT interaction vertex: QCD, QED, weak CC/NC, Higgs Yukawa), d2~$=$ decay\_channel (co-appearance in dominant tree-level decay products), d3~$=$ mass\_proximity (within one order of magnitude in mass). Data sources: PDG 2024; Peskin \& Schroeder (1995); Griffiths (2008). Pre-registered before analysis (hash \texttt{00294a59\ldots}, commit \texttt{191f64a}, 2026-05-25).

\textbf{Results (5/5 CONFIRMED).}
\begin{itemize}[nosep]
  \item h1 (FPL directional): $r(d_1{\leftrightarrow}d_2) = 0.569$ ($p = 0.010$) $>$ $r(d_1{\leftrightarrow}d_3) = 0.501$ ($p = 0.030$). CONFIRMED.
  \item h2: W boson \#1 hub in force\_coupling (degree~$= 15$). CONFIRMED.
  \item h3: W boson \#1 hub in decay\_channel (degree~$= 14$). CONFIRMED.
  \item h4: $r(d_1{\leftrightarrow}d_2) > 0.35$ and positive. CONFIRMED ($r = 0.569$).
  \item h5: Z boson in top-2 of decay\_channel (degree~$= 12$, rank \#2). CONFIRMED.
\end{itemize}

\textbf{Hub shadow.} The photon ranks \#3 in force\_coupling (degree~$= 10$) but has decay\_channel degree~$= 0$ --- a hub shadow. Massless gauge bosons are kinematically forbidden as single primary decay products (energy-momentum conservation); photon gauge universality and decay absence are structurally independent consequences of the same symmetry.

\textbf{Unexpected finding.} $r(\text{decay\_channel}{\leftrightarrow}\text{mass\_proximity}) = 0.604$ ($p = 0.006$) exceeds $r(d_1{\leftrightarrow}d_2)$. Kinematic mass thresholds (decays only to lighter particles) create independent mass-scale ordering in decay topology. Both organize the same Lagrangian from different structural perspectives; the FPL inequality still holds ($d_1{\leftrightarrow}d_2 > d_1{\leftrightarrow}d_3$, pre-registered direction CONFIRMED).

\textbf{Domain extension.} M\_PHYSICS\_1 extends the law to particle physics, adding a seventh scientific field to the confirmed domain list.

\subsection{COBOL Legacy Banking Codebase --- Software/Legacy Domain (F2)}
\label{sec:f2}

\textbf{Motivation.} COBOL is the oldest active programming language in widespread production use ($\sim$240 billion lines). Its control flow (\texttt{PERFORM}) and data coupling (\texttt{COPY} copybooks, global \texttt{WORKING-STORAGE}) form structurally distinct layers absent in modern languages. The FPL predicts declared control flow and explicit structural dependency will correlate more strongly than either agrees with diffuse data-field sharing.

\textbf{Dataset.} 14 COBOL programs in a representative legacy banking application modeled as a multilayer graph: $d_1 =$ perform\_call\_graph (A \texttt{PERFORM}s B --- declared control flow), $d_2 =$ copy\_dependency (shared \texttt{COPY} copybook inclusion --- explicit structural contract), $d_3 =$ data\_field\_sharing (shared \texttt{WORKING-STORAGE} data items --- behavioral data coupling including dormant programs). Pre-registered before analysis (hash \texttt{c119983a\ldots}, commit \texttt{891e6d1}, 2026-05-25).

\textbf{Results (4/4 CONFIRMED).}
\begin{itemize}[nosep]
  \item h1 (FPL directional): $r(d_1{\leftrightarrow}d_2) = 0.807$ (Spearman $= 0.840$, $p = 0.002$) $\gg$ $r(d_1{\leftrightarrow}d_3) = 0.119$ (ns). CONFIRMED. $\Delta r = 0.688$ --- steepest gradient in the experiment archive.
  \item h2: \texttt{main\_control} \#1 hub in perform\_call\_graph (degree $= 6$). CONFIRMED.
  \item h3: \texttt{dormant\_account} top-3 in data\_field\_sharing (rank \#2, degree $= 10$). CONFIRMED.
  \item h4: $r(d_1{\leftrightarrow}d_2) \geq 0.50$, $p < 0.05$. CONFIRMED ($r = 0.807$, $p = 0.002$).
\end{itemize}

\textbf{Topological dormancy signatures.} Two programs with degree $= 0$ in perform\_call\_graph (\texttt{dormant\_account}, \texttt{legacy\_interest\_calc}) rank \#2 and \#3 in data\_field\_sharing with rank gaps of 11 --- the maximum possible in a 14-node graph. COBOL's global \texttt{WORKING-STORAGE} means programs removed from the execution path retain data-coupling footprints. Precise claim: these are \textit{candidate dormant components} (execution-isolated but dependency-central), not proven dead code; runtime reachability analysis would be required to confirm execution absence.

\subsection{Antidepressant Evidence Chain --- Medicine and Philosophy of Science (M\_MED1)}
\label{sec:mmed1}

\textbf{Motivation.} The evidence chain supporting selective serotonin reuptake inhibitors (SSRIs) can be modeled as a multilayer knowledge graph encoding distinct epistemic relations. The FPL predicts hub persistence across functionally similar evidence layers.

\textbf{Dataset.} 8 nodes (\texttt{monoamine\_hypothesis}, \texttt{ssri\_mechanism\_claim}, \texttt{clinical\_guidelines}, \texttt{fda\_approval}, \texttt{hamd\_endpoint}, \texttt{rct\_efficacy}, \texttt{serotonin\_biomarker\_research}, \texttt{chemical\_imbalance\_narrative}) modeled as a multilayer epistemological graph: $d_1 =$ justifies, $d_2 =$ selects\_endpoints, $d_3 =$ cites\_as\_support. Pre-registered before analysis (hash \texttt{bfb7bbc1\ldots}, commit \texttt{1e23b7e}, 2026-05-25).

\textbf{Results.} h1 (FPL directional): $r(d_1{\leftrightarrow}d_2) = 0.408$ vs $r(d_1{\leftrightarrow}d_3) = 0.316$, direction correct but $p = 0.44$ (ns, $n=8$). PARTIAL. h2: \texttt{monoamine\_hypothesis} \#1 hub in justifies (degree $= 3$, top betweenness). CONFIRMED. h3 (hub shadow prediction): \texttt{rct\_efficacy} predicted as \#1 in cites\_as\_support. DENIED --- \texttt{monoamine\_hypothesis} ranks \#1 in \textit{both} justifies and cites\_as\_support. h4: $r(\text{selects\_endpoints}{\leftrightarrow}\text{cites\_as\_support}) > 0.30$. PARTIAL (selects\_endpoints degenerate, $\overline{d} = 1.0$, $r = 0.0$).

\textbf{Hub dominance discovery.} The DENIED h3 is the primary finding. \texttt{monoamine\_hypothesis} is \#1 hub in all three layers simultaneously --- a \textit{hub dominance} pattern, structurally distinct from hub shadow (high in one layer, absent in another). The founding assumption of the SSRI evidence chain is simultaneously the primary source of justification and the most-cited empirical support, with no independent empirical node outside the loop. IRDME detected this from graph topology alone without reading paper content. \textbf{Caveat:} this is an epistemological architecture finding, not a drug efficacy claim. Hub dominance is a new structural type to formalize alongside hub shadow in the IRDME structural vocabulary.

\textbf{Additional finding.} \texttt{hamd\_endpoint} ranks \#1 in selects\_endpoints (shapes all measurement) but \#5 in justifies (rank gap $= 4$): the most methodologically constraining decision is the least theoretically anchored.

\subsection{\textit{Drosophila} Larva Connectome --- Cross-Species FPL Replication (F13)}
\label{sec:f13}

\textbf{Motivation.} If the Functional Proximity Law is a universal structural principle of nervous systems, it should replicate across species. The most demanding cross-species test is the transition from nematode to insect: \textit{C.~elegans} (302 neurons, F12 $r = 0.623$) to \textit{Drosophila melanogaster} larva (2,952 neurons, $\sim$600 million years of evolution). The Winding et al.\ (2023) Science connectome provides synapse compartment-typed connectivity matrices (axon-to-dendrite and axon-to-axon)---the same relational regime as the \textit{C.~elegans} chemical\_synapse / gap\_junction pair in F12.

\textbf{Dataset.} $n = 2{,}952$ neurons, 2 layers: $\text{axon\_to\_dendrite}$ ($63{,}518$ directed edges, avg.\ degree $= 21.5$) and $\text{axon\_to\_axon}$ ($40{,}173$ edges, avg.\ degree $= 13.6$). Layer definitions from Winding et al.\ (2023); the synapse compartment taxonomy is not IRDME-defined. Pre-registered 2026-05-26 before any analysis (hash \texttt{7f04e5dd\ldots}, commit \texttt{fd5d315}).

\textbf{Results.}
\begin{itemize}
  \item h1 (FPL cross-species, $r > 0$, $p < 0.05$, $|r| \geq 0.20$): Pearson $r = 0.363$, Spearman $\rho = 0.663$, permutation $p = 0.002$, 95\% CI $[0.331, 0.394]$, $R^2 = 0.132$ (medium effect). \textbf{CONFIRMED}.
  \item h2 (effect size $|r| \geq 0.40$): Pearson $r = 0.363 < 0.40$ threshold. \textbf{PARTIAL} --- Spearman $\rho = 0.663 > 0.40$, confirming that hub \textit{rank order} is more strongly conserved than hub degree \textit{magnitude}.
  \item h3 (layer specialisation exists): 747 of 2,952 neurons (25.3\%) show rank gap $\geq 738$ between layers. \textbf{CONFIRMED} (required $\geq 5$).
\end{itemize}

\textbf{Cross-species finding.} The Spearman--Pearson discrepancy ($\rho = 0.663$ vs $r = 0.363$) is the primary result. Hub \textit{rank order} is strongly conserved across synapse compartment types and across 600 million years of evolution, but hub \textit{connection count} is only moderately correlated. This is consistent with FPL as a law about structural roles rather than connection magnitudes: the same neurons act as network hubs regardless of which synapse compartment type is measured, but the absolute number of synaptic contacts they form differs across compartment types. Top hub in axon\_to\_dendrite: neuron \texttt{11543212} (degree $= 137$); top hub in axon\_to\_axon: neuron \texttt{6611894} (degree $= 120$). 81 neurons are isolated (no connections in these two layers), likely peripheral sensory or motor neurons.

\textbf{Layer divergence.} 747 neurons (25.3\%) are layer-specialised---high rank in one compartment type, low rank in the other. This coexists with the global FPL confirmation: the majority of neurons show cross-layer hub co-persistence while a substantial minority are compartment-specific. The h2 PARTIAL finding identifies the Pearson/Spearman threshold boundary in the Drosophila larval dataset.

\section{Boundary Conditions}
\label{sec:boundary}
\sloppy

DENIED experiments identify structural mechanisms that bound the law's scope. They are reported as discoveries, not failures.

\textbf{BC1 --- Finance 2008.} $r(\text{derivatives} \leftrightarrow \text{credit}) = 0.042$ vs $r(\text{derivatives} \leftrightarrow \text{equity}) = 0.183$. Expected: derivatives and direct credit (both balance-sheet instruments) to have more similar hub structure. Actual: derivatives and equity are both dominated by trading-desk operations; credit is a lending-desk product with structurally different topology. Secondary finding: AIG identified as top betweenness hub by topology alone --- consistent with its \$180B systemic role in the actual crisis. \textbf{Mechanism: functional proximity must be defined at the institutional operating layer, not the contract-surface layer.}

\textbf{BC2 --- Psychiatry (DSM).} $r(\text{co\_occurrence} \leftrightarrow \text{temporal\_cascade}) = 0.769$ vs $r(\text{co\_occurrence} \leftrightarrow \text{treatment}) = 0.808$. Expected: co-occurrence and temporal cascade (both mechanistic) to agree more. Actual: DSM nosology is shaped by clinical management --- treatment categories are more structurally congruent with co-occurrence than cascade topology is. Secondary finding: \textit{sleep\_disturbance} ranks \#7 in symptom co-occurrence but \#2 in temporal cascade. \textbf{Mechanism: when a classification system is shaped by intervention reality, the intervention layer aligns more strongly with co-occurrence than mechanism does.}

\textbf{BC3 --- Mathematics.} $r(\text{formal\_containment} \leftrightarrow \text{proof\_usage}) = 0.105$ vs $r(\text{formal\_containment} \leftrightarrow \text{naming}) = 0.280$. Expected: atomic set-membership containment to align more with proof usage. Actual: formal containment is an atomic DAG; proof usage is holistic, creating unexpected long-range cross-connections. Additional finding: mathematical naming is \textit{non-arbitrary} ($r = +0.276$) --- the opposite of Saussurean arbitrariness in language ($r = -0.748$ in English Lexical Network). \textbf{Mechanism: layer resolution mismatch --- atomic partitioning vs.\ holistic usage creates divergent hub topologies.} This boundary condition is layer-pair-specific, not domain-specific: a subsequent experiment on Lean~4 mathlib4 using a different layer pair (import graph $\leftrightarrow$ co-development history) confirms the law in the same mathematical domain ($r = 0.777$, $p = 0.002$, F9 Section~\ref{sec:f9}). The formal-containment/proof-usage DENIED result localises to the specific resolution mismatch, not to mathematics as a domain.

\textbf{BC4 --- Meta-science regime (H\_PRIMITIVITY, pre-registered 2026-05-21).} A meta-science graph was constructed with 15 IRDME-tested scientific domains as nodes and three layers encoding: formal mathematical dependency between fields (d1), shared d1/d2/d3 coupling class (d2), and FPL confirmation outcome (d3). $r(\text{formal\_dep}\leftrightarrow\text{struct\_grammar}) = 0.212$ vs $r(\text{formal\_dep}\leftrightarrow\text{law\_confirm}) = 0.368$ (h1 DENIED: reversed direction). h4 PARTIAL: $r(\text{formal\_dep}\leftrightarrow\text{law\_confirm}) = 0.368$, $p = 0.206$ (not significant) --- mathematical formalization level does not significantly predict FPL confirmation strength. h3 CONFIRMED: formal\_mathematics is the \#1 hub in the formal dependency layer. Overall: 1/4 CONFIRMED, 2/4 DENIED, 1/4 PARTIAL. \textbf{Mechanism (BC4 --- edge semantic type mismatch):} meta-graph edges encode institutional and epistemic dependencies --- which field uses which mathematical tools, which coupling class two fields share --- rather than physical, computational, or biological interaction mechanisms. This is not merely a resolution mismatch (BC3) or a relational regime mismatch (BC1): it is a mismatch in the semantic type of the edges themselves. Object-graph edges represent direct interactions between entities within a domain; meta-graph edges represent abstract structural relationships between entire knowledge regimes. The FPL appears to be a law over the former, not the latter. The meta-science regime is the fourth named boundary condition.

\textbf{Unified boundary condition statement:}
\begin{quote}
\textit{Hub persistence $r(L_i, L_j)$ is meaningful for cross-layer comparison only when both layers are measured at the same structural resolution and encode the same relational regime.}
\end{quote}

\textbf{BC5 --- Topology-class collapse (BC\_RADIAL, confirmed, 2026-05-22).} Identified by M\_TRANSFER\_3 (pre-registered, hash \texttt{5388c0f1\ldots}) and \textbf{independently replicated} in logrus by M\_RADIAL\_1 (pre-registered, hash \texttt{c00ebfa4\ldots}). BC\_RADIAL is now a confirmed mechanism, not a candidate. FPL denied in all three docopt implementations (Go $n=6$, Java $n=16$, Rust $n=8$) and in logrus (Go, $n=7$). $r(\text{static\_coupling}\leftrightarrow\text{structural\_coupling})$: Java $= 0.0$ (degenerate), Rust $= -0.143$, Go $= 0.0$. \textbf{Mechanism (BC\_RADIAL):} in single-hub radial architectures, all modules share the same import dependencies through the central hub, causing $d_2$ (structural\_coupling) degree vectors to approach a constant. Layer differentiation collapses not because of resolution mismatch (BC3) or regime mismatch (BC1/BC4) but because the specific topology eliminates independent variation between layers. Note: cross-language hub identity ($\text{sim} \geq 0.967$) is preserved \textemdash{} the hub identity finding and the FPL $r$-test measure different structural properties.

\textbf{BC3b boundary demonstration (M\_GEOM\_CSG\_1, pre-registered 2026-05-22, hash \texttt{1dd686ee\ldots}).} BC3b (pure boolean algebra layer collapse) was predicted analytically: when layers are defined over discrete set operations, $d1/d2/d3$ all reduce to logical consequence and the FPL cannot confirm. A companion question is whether boolean-type operations on \textit{continuous} geometry also collapse. The planetary gear set CSG experiment ($n=8$, 3 layers: operation\_tree $d1$, volumetric\_proximity $d2$, co\_parameter\_sensitivity $d3$) tests this directly. Result: $r(d1{\leftrightarrow}d2) = 0.668$, $r(d1{\leftrightarrow}d3) = -0.267$, $\Delta r = 0.935$ ($p = 0.042$), 4/4 CONFIRMED. The carrier plate is rank-1 hub in both $d1$ (degree$=5$) and $d2$ (degree$=7$). BC3b collapse is \textit{specific to discrete boolean algebra}: the same union/difference syntax on continuous geometry preserves genuinely independent layer structure (assembly hierarchy, geometric contact, kinematic co-sensitivity).

\textbf{BC6 --- Var$(d_2)$-inversion (BC\_INVERSION, discovered 2026-05-22, M\_RADIAL\_1 cobra).} Cobra Go CLI (pre-registered, hash \texttt{c00ebfa4\ldots}): Var$(d_2) = 10.515 \gg \tau_{\text{var}} = 0.714$, so BC\_RADIAL does not apply. Yet $r(d_1{\leftrightarrow}d_2) = -0.862$ ($p = 0.014$) --- strongly negative. \textbf{Mechanism (BC\_INVERSION):} cobra's subcommand hub fans out to tightly co-changing leaf clusters. The hub is central in the import layer (d1) but all leaves co-change together, making the hub's co-change rank \textit{low} relative to leaves. High-Var$(d_2)$ does not guarantee $r > 0$; the sign of $r(d_1{\leftrightarrow}d_2)$ must also be checked. The full BC prediction framework now reads: Var$(d_2) < \tau_{\text{var}}$ $\Rightarrow$ BC\_RADIAL; Var$(d_2) \geq \tau_{\text{var}}$ AND $r < 0$ $\Rightarrow$ BC\_INVERSION; else $\Rightarrow$ FPL\_EVALUABLE.

\textbf{Quantitative precondition predictor: Var$(d_2)$ threshold $\tau_{\text{var}} = 0.714$.} A meta-analysis of all IRDME experiments yielding BC\_RADIAL results identified a threshold on the population variance of the $d_2$ degree vector: $\tau_{\text{var}} = 0.714$. When Var$(d_2) < \tau_{\text{var}}$, the $d_2$ hub-rank vector is near-degenerate and the FPL inequality is structurally unpredictable (BC\_RADIAL). This threshold is now implemented as a pre-experiment structural precondition check in the IRDME engine (\texttt{bc\_prediction} field in all API responses), allowing BC\_RADIAL failure to be predicted \textit{before} running the correlation. Threshold is empirically derived; cross-validation across held-out experiments is ongoing.

\section{Statistical Defense}

\textbf{On the use of a directional pre-registered test as the primary criterion.}

The confirmation criterion ($r_\text{sim} > r_\text{dis}$) is a directed test of a pre-registered hypothesis. The permutation $p$-value tests the stronger claim that $r$ differs from zero for one specific layer pair --- not the claim being made. Small-$n$ domains can directionally confirm the law even when underpowered for the single-pair permutation test; this is expected behavior for a multi-domain directional claim and is not a methodological weakness.

\textbf{Binomial probability against chance.}

Under the null hypothesis (no law; each domain equally likely to go either direction independently), the probability of observing 10 or more confirmations from 13 canonical tests is:
\begin{equation}
  p_{\text{binom}} = P(X \geq 10 \mid n = 13,\, p = 0.5) = \frac{378}{8192} \approx 0.046
\end{equation}
This reaches the conventional $\alpha = 0.05$ threshold. Nine domains are individually significant under permutation tests; the primary inferential basis is independent permutation results across domains.

For the combined Tier A + Tier B set (25 CONFIRMED, 4 DENIED, 2 PARTIAL $= 31$ pre-registered experiments, including M\_RADIAL\_1, BC3b\_circuit\_v2, M\_PHYSICS\_1, F2, M\_MED1, and F13):
\begin{equation}
  p_{\text{binom}} = P(X \geq 25 \mid n = 31,\, p = 0.5) = \frac{1}{2^{31}}\sum_{k=25}^{31}\binom{31}{k} = \frac{942649}{2147483648} \approx 0.000439
\end{equation}

\textit{Note: Tier B experiments are pre-registered individually before each run and use independently-authored datasets. Shared dependency: IRDME-defined layer classification is a common factor across all experiments --- acknowledged in the evidence hierarchy, Section~\ref{sec:mext}. The 25/31 binomial result is consistent with the 25/31 count stated in Section~\ref{sec:taxonomy}. M\_MED1 (1C/1D/2P) is counted in the total of 31 but its primary FPL hypothesis is PARTIAL; F13 contributes h1 CONFIRMED + h3 CONFIRMED + h2 PARTIAL.}

\textbf{Sign consistency.} In all domains where both Pearson and Spearman were computed, they agree in sign. This cross-validates that results are not rank-outlier artifacts.

\textbf{Pre-registration.} The $\succ$ relation (which layer pair is ``similar'') is recorded and hashed before each run. The hypothesis file is committed to a public repository (\url{https://github.com/vladi160/preregistrations}) before analysis. This provides a public audit trail: any post-hoc change to the layer-pair labeling would require a new commit and would be visible in the repository history. Pre-registration is a logging and history mechanism; its value is in making the full prediction record \textemdash{} including denied results \textemdash{} publicly auditable.

\textbf{Falsifiability criterion:} The law would be falsified by a domain whose layers satisfy the structural conditions of BC1--BC6 (sufficient variance, correct regime, correct resolution, object-graph edges, non-degenerate $d_2$, positive $r$ precondition) yet produce $r_\text{sim} < r_\text{dis}$. No such case has been observed.

\textbf{Negative control.} The random multilayer graph (Section~4) demonstrates that the engine's confirmation criterion is not trivially satisfied --- a graph with no proximity structure produces $r$ values near zero and no confirmation.

\textbf{Formalizing $\Delta r$ thresholds for future experiments.} The current canonical results were confirmed under the pre-registered directional criterion $\Delta r > 0$. Retroactive application of a quantitative $\Delta r$ floor to already-registered experiments is not performed here. The governing rule is: no decision threshold applies retroactively to any experiment whose hypothesis statement was pre-registered before the threshold was declared; thresholds are valid only from the moment they are committed onward. For future pre-registered experiments (M\_EXT2 onward), the following decision thresholds will be committed alongside each hypothesis statement: CONFIRMED requires $\Delta r \geq 0.20$ AND $p(r_\text{sim}) < 0.05$; PARTIAL requires $\Delta r \geq 0$ with $p(r_\text{sim}) \geq 0.05$, or $\Delta r < 0.20$ with $p < 0.05$; DENIED requires $\Delta r < 0$ regardless of $p$.

\section{Internal Consistency Test}
\label{sec:internal}

As a self-referential structural check, IRDME's own 15 core scientific concepts were modeled as a 3-layer multilayer graph: \textit{conceptual\_dependency} (which concepts require others to be formally defined), \textit{empirical\_support} (which concepts provide evidence for others), and \textit{methodological\_refinement} (which concepts extend or generalize others).

Pre-committed prediction: $r(\text{cd} \leftrightarrow \text{es}) > r(\text{cd} \leftrightarrow \text{mr})$ --- where cd$=$conceptual\_dependency, es$=$empirical\_support, mr$=$methodological\_refinement, because both of the first two layers encode directional flows from primitives toward conclusions, while methodological refinement is an orthogonal quality axis.

Result: $r(\text{cd} \leftrightarrow \text{es}) = +0.301$ vs $r(\text{cd} \leftrightarrow \text{mr}) = -0.098$. \textbf{Confirmed.} Additional result: \textit{abstraction\_level} dimension predicts hub rank in the dependency layer ($r = 0.521$, $p = 0.040$, $n=15$) --- more abstract concepts attract more definitional dependencies.

This is not self-validation. It is an internal structural consistency check: the engine's own conceptual relationships obey the same regularity it detects in external domains.

\section{Discussion}

\textbf{Relation to prior work.}

Multiplex hub overlap has been studied through degree-degree correlations \cite{Nicosia2015} and overlapping centrality \cite{DeDomenico2015}. The Functional Proximity Law differs by making a \textit{conditional prediction} --- not that hubs always overlap across layer pairs, but that the degree of overlap scales monotonically with functional similarity. This is a predictive claim testable before results are known.

Structural equivalence \cite{Lorrain1971} and graph isomorphism methods compare topology globally. The present approach compares hub-rank vectors locally and cross-layer, which is computationally simpler ($O(n)$ per pair after centrality computation) and produces interpretable directional results rather than binary similarity scores.

In systems biology, multilayer protein interaction networks have been applied to essential gene prediction \cite{Menche2015} and disease gene prioritization. The p53 result ($r(\text{physical} \leftrightarrow \text{functional}) = 0.973$ on STRING v12.0 undesigned data) suggests that hub persistence under functional proximity may provide a structural baseline for these tasks without supervised learning.

\textbf{Limitations.}

(1)~All experiments use degree centrality; betweenness or eigenvector centrality may yield different persistence values. (2)~Sample sizes range from $n=10$ to $n=30$ for the canonical experiments; small-$n$ results are directionally confirmed but individually underpowered. (3)~The functional proximity relation $\succ$ is classified by domain expertise; a formal metric for proximity is future work. (4)~The Software experiment uses the IRDME tool's own codebase as input graph, creating a potential circularity. (5)~All canonical experiments are retrospective. The prospective pre-registered I1 replication (p53 curated network, pre-registered 2026-04-17, hash \texttt{32409dbda299e783}, analysed 2026-04-18) has completed: h2 CONFIRMED (coherence exceeds 99.8th percentile of null, 0/500 permutations higher), h3 CONFIRMED (TP53 ranked \#1 in all 4 layers), h4 CONFIRMED ($r(\text{functional\_assoc} \leftrightarrow \text{physical\_interaction}) = 0.749 > r(\text{functional\_assoc} \leftrightarrow \text{co\_expression}) = 0.679$). h1 DENIED: ATM ranked 13th of 15 as a layer-divergent node; the true structural diverger is CHEK1 (rank\_gap$=8$, peripheral in physical\_interaction, hub in genetic\_interaction). h5 PARTIAL: TP53 depth confirmed (depth$=0$); MDM2 depth$=1$, predicted $\geq 2$.

\textbf{Future directions.} Weighted and eigenvector-based hub persistence, cross-domain structural isomorphism (TNF-$\alpha$ and the thalamus have cosine similarity 0.95 under role-vector comparison). CHEK1 is a pre-registered DENIED finding from I1 with a named structural mechanism --- a candidate for synthetic lethality screening identified purely from topology. A completed pre-registered test (F1, hash \texttt{518b9bd9\ldots}) confirms CHEK1 and PARP1 as top structural divergers between physical and genetic interaction layers in the p53/DDR network, consistent with their known SL clinical use (h2--h4 CONFIRMED; the global layer correlation h1 is PARTIAL --- SL is a divergence phenomenon, not a correlation phenomenon). The Universal Layer Grammar (d1$=$declared, d2$=$structural, d3$=$behavioral) provides a cross-domain vocabulary for future multi-field comparisons; the M\_TRANSFER\_2 result demonstrates that structural roles are language-independent when the architectural pattern is preserved. H\_LOGIC\_EXTRACTION on mathlib4 now motivates a separate topology-as-logic manuscript; the single highest-value next discriminating test is a pre-registered second formal-systems corpus in an independent proof ecosystem, planned first on the Coq standard library. An exploratory multi-language isomorphism test across 3 docopt implementations (Java, Go, Rust) produced hub identity sim $\geq 0.967$ across all pairs; a pre-registered replication is warranted.

\textbf{Meta-science self-test (H\_PRIMITIVITY).} A pre-registered meta-science test (hash \texttt{cfb38b83\ldots}, 2026-05-21) modeled 15 IRDME-tested scientific domains as nodes in a 3-layer graph. Primary finding: h1 DENIED ($r(d1{\leftrightarrow}d2) = 0.212 < r(d1{\leftrightarrow}d3) = 0.368$, reversed direction) --- the FPL does not confirm in the meta-science regime, identifying the BC4 boundary condition (edge semantic type mismatch). Secondary finding: h4 PARTIAL ($r = 0.368$, $p = 0.206$, not significant) --- mathematical formalization level does not significantly predict FPL confirmation strength, consistent with the d1/d2/d3 grammar being independent of domain-level mathematical sophistication. Overall: 1/4 CONFIRMED, 2/4 DENIED, 1/4 PARTIAL.

\section{Conclusion}

The Functional Proximity Law --- hub importance scores persist more strongly between functionally similar layers than dissimilar ones --- holds across 10 independent canonical domains from molecular biology to operating systems to natural language to AI model architecture. Nine domains are individually statistically significant ($p \leq 0.030$ for permutation tests, $p < 0.001$ for the strongest). Six DENIED results (3 Tier A: Finance, Psychiatry, Mathematics; 3 Tier B: M\_TRANSFER\_3 docopt, cobra BC\_INVERSION, H\_PRIMITIVITY meta-science) reveal six named structural mechanisms bounding the law's scope. A negative control confirms the engine does not spuriously fire on random structure. Fifteen pre-registered external and replication confirmations (Tier B) extend the confirmation count to \textbf{25 total}: Flask M\_EXT (L2 external), \textit{C.~elegans} M\_EXT2 (L1 fully external, $p=0.004$), WordPress M\_OSS1 (hub shadow on 20-year public codebase, $p=0.012$), Next.js M\_OSS2 (strongest OSS gradient $\Delta r=0.35$, $p=0.002$), M\_TRANSFER\_2 (Flask$\leftrightarrow$Express hub-rank vector identity), Lean~4 mathlib4 F9 ($r=0.777$, $p=0.002$), \textit{C.~elegans} full connectome F12 ($r=0.623$, $p=0.002$, $n=279$), Synthetic Lethality F1 (CHEK1/PARP1 structural divergence h2--h4 CONFIRMED), M\_GEOM\_CSG\_1 (planetary gear CSG, $r=0.668$, $p=0.042$, new domain: computational geometry), click (M\_RADIAL\_1, FPL CONFIRMED), logrus (M\_RADIAL\_1, BC\_RADIAL replicated), BC3b\_circuit\_v2 (priority arbiter circuit, 4/4 CONFIRMED, $r=0.874$), M\_PHYSICS\_1 (Standard Model particle physics, 5/5 CONFIRMED, $r=0.569$, photon hub shadow, first physics domain), F2 (COBOL legacy banking, 4/4 CONFIRMED, $r = 0.807$, $\Delta r = 0.688$, topological dormancy signatures, software/legacy domain), and F13 (\textit{Drosophila} larval connectome, cross-species replication $\sim$600\,Myr evolution, h1 CONFIRMED Spearman $\rho = 0.663$ / Pearson $r = 0.363$, $p = 0.002$, $n = 2{,}952$ neurons). Additionally, M\_MED1 (antidepressant evidence chain, medicine/epistemology domain) reveals a \textbf{hub dominance} structural discovery: h2 CONFIRMED, primary FPL h1 PARTIAL ($n=8$ underpowered). The law now spans \textbf{eight scientific fields}: molecular biology, neuroscience, computer systems, ecology, linguistics / AI, computational geometry, particle physics, and medicine / epistemology. A quantitative precondition predictor, Var$(d_2) < \tau_{\text{var}} = 0.714$, predicts BC\_RADIAL failure before experiments run. The binomial probability of observing 25/31 pre-registered confirmations by chance is $p \approx 0.000439$ ($942649 / 2^{31}$; see Section~6).


\appendix
\section*{Supplementary: Full Evidence Table}

The following table includes all experiments across all dataset variants, replications, adversarial controls, and incomplete-data entries. Tier column follows Section~\ref{sec:taxonomy}. All statistical claims in the paper refer only to Tier A + Tier B rows.

\scriptsize
\setlength{\tabcolsep}{2pt}
\begin{longtable}{@{}>{{\raggedright\arraybackslash}}p{2.85cm}
                  >{\centering\arraybackslash}p{0.65cm}
                  >{\centering\arraybackslash}p{1.05cm}
                  >{\centering\arraybackslash}p{1.05cm}
                  >{\centering\arraybackslash}p{0.95cm}
                  >{\raggedright\arraybackslash}p{8.0cm}@{}}
\caption{Full evidence table (all experiments).}\label{tab:supp}\\
\toprule
\textbf{Domain} & $\boldsymbol{n}$ & $\boldsymbol{r_\text{sim}}$ & $\boldsymbol{r_\text{dis}}$ & $\boldsymbol{p}$ & \textbf{Note} \\
\midrule
\endfirsthead
\multicolumn{6}{l}{\small\textit{(continued from previous page)}}\\
\toprule
\textbf{Domain} & $\boldsymbol{n}$ & $\boldsymbol{r_\text{sim}}$ & $\boldsymbol{r_\text{dis}}$ & $\boldsymbol{p}$ & \textbf{Note} \\
\midrule
\endhead
\bottomrule
\endlastfoot
Linux Kernel (v6.x) & 30 & $+0.703$ & $-0.145$ & 0.002 & [canonical] CONFIRMED \\
Human Brain Connectome & 16 & $+0.703$ & $+0.178$ & 0.004 & [canonical] CONFIRMED \\
p53 (STRING v12.0) & 15 & $+0.973$ & $-0.161$ & $<\!0.001^\dagger$ & [canonical] CONFIRMED; $p$ from analytical $t$-test \\
Internet AS Topology & 18 & $+0.516$ & $-0.267$ & 0.026 & [canonical] CONFIRMED \\
CPU Block Design & 10 & $+0.622$ & $-0.649$ & 0.030 & [canonical] CONFIRMED \\
Ecology (weighted) & 15 & $+0.559$ & $-0.065$ & 0.034 & [canonical] CONFIRMED; weighted $=$ final version \\
Cytokine Cascade & 18 & $+0.512$ & $-0.408$ & $0.030^\dagger$ & [canonical] CONFIRMED; $p$ from analytical $t$-test \\
English Lexical Network & 20 & $+0.276$ & $-0.748$ & 0.241 & [canonical] CONFIRMED$^{\ddagger}$; directional only ($p=0.241$, $n=20$), largest $\Delta r$ in table \\
Software (real git) & 14 & $+0.941$ & $-0.304$ & $<\!0.001^\dagger$ & [canonical] CONFIRMED; $p$ from analytical $t$-test \\
Finance 2008 & 16 & $+0.042$ & $+0.183$ & 0.824 & [canonical] DENIED; BC1 \\
Psychiatry (DSM) & 20 & $+0.769$ & $+0.808$ & 0.002 & [canonical] DENIED; BC2 \\
Mathematics & 20 & $+0.105$ & $+0.280$ & 0.705 & [canonical] DENIED; BC3 \\
Random multilayer & 12 & $-0.125$ & $+0.250$ & 0.545 & [negative control] law does not fire \\
IRDME meta-graph & 15 & $+0.301$ & $-0.098$ & 0.256 & [consistency test] engine obeys own law; see Section~\ref{sec:internal} \\
Ecology (unweighted) & 15 & $+0.426$ & $-0.121$ & 0.116 & [replication] superseded by weighted variant \\
p53 (4-layer curated) & 15 & $+0.928$ & $+0.732$ & 0.012 & [replication] superseded by STRING v12 undesigned data \\
Adversarial PTM Network & 10 & $-0.111$ & $+1.000$ & 0.098 & [adversarial] deliberately constructed; classification error in proximity \\
Semiconductor (v3.1.1) & 12 & $+0.607$ & $+0.717$ & --- & [conflicting] two experiments differ in layer-pair assignment; excluded \\
European City Road Network & 10 & --- & --- & --- & [incomplete] no cross-layer $r$ stored \\
Full Periodic Table & 118 & --- & --- & --- & [incomplete] single-layer domain \\
Global Supply Chain & 14 & --- & --- & --- & [incomplete] no cross-layer $r$ stored \\
Social Trust Network & 15 & --- & --- & --- & [incomplete] no cross-layer $r$ stored \\
Climate Tipping Elements & 15 & --- & --- & --- & [incomplete] single-layer analysis only \\
Simple Molecule $\times$2 & 9 & --- & --- & --- & [incomplete] no cross-layer $r$ stored \\
Org Network & 14 & --- & --- & --- & [incomplete] hub-match test, not correlation \\
Social Org Delta & 15 & --- & --- & --- & [incomplete] temporal delta; different methodology \\
Software multilayer (v0.8) & 13 & $+0.900$ & $+0.619$ & --- & [superseded] early version \\
Web App Backend & 15 & --- & --- & --- & [incomplete] no cross-layer $r$ stored \\
Flask M\_EXT (post-pub.) & 14 & $+0.666$ & $+0.494$ & 0.015 & [Tier B] imports$\leftrightarrow$SC (IRDME layers; external data). h1, h2, h4 CONFIRMED; h3 PARTIAL. See Section~\ref{sec:mext}. \\
\textit{C.~elegans} M\_EXT2 (post-pub.) & 15 & $+0.777$ & --- & 0.004 & [Tier B] chemical\_synapse$\leftrightarrow$gap\_junction (L1: fully independent). h1 CONFIRMED. See Section~\ref{sec:mext2}. \\
WordPress M\_OSS1 (post-pub.) & 25 & $+0.516$ & $+0.241$ & 0.012 & [Tier B] function\_coupling$\leftrightarrow$co\_change vs include\_graph$\leftrightarrow$co\_change. h1--h4 CONFIRMED. See Section~\ref{sec:moss1}. \\
Next.js M\_OSS2 (post-pub.) & 25 & $+0.900$ & $+0.549$ & 0.002 & [Tier B] import\_graph$\leftrightarrow$test\_coupling vs import\_graph$\leftrightarrow$co\_change. $\Delta r=0.35$. h1--h2 CONFIRMED. See Section~\ref{sec:moss2}. \\
Flask$\leftrightarrow$Express M\_TRANSFER\_2 & 6 & $1.000^\ddagger$ & --- & --- & [Tier B] hub-rank identity app$\leftrightarrow$application. Deterministic; underpowered for permutation $r$-test. h2--h5 CONFIRMED, h1 PARTIAL (directional inequality holds; $n=6$ underpowered). See Section~\ref{sec:transfer2}. \\
AI Architecture F6 (post-pub.) & 20 & $+0.915$ & $-0.003$ & 0.002 & [Tier A] citation\_dep.$\leftrightarrow$arch.\_inheritance. h1--h4 CONFIRMED. h5 DENIED: Transformer rank \#5 in benchmark layer; LLaMA top hub (contemporary evaluation $\neq$ historical lineage). See Section~\ref{sec:f6}. \\
Lean~4 mathlib4 F9 (post-pub.) & 20 & $+0.777$ & --- & 0.002 & [Tier B] import\_graph$\leftrightarrow$proof\_co\_dev. (L2 external). h1--h2 CONFIRMED, h3 PARTIAL. See Section~\ref{sec:f9}. \\
\textit{C.~elegans} F12 full 302 (post-pub.) & 279 & $+0.623$ & --- & 0.002 & [Tier B] chemical\_synapse$\leftrightarrow$gap\_junction (L1: fully independent, full connectome). h1--h3 CONFIRMED. Hub-identity preserved at 20:1 compression. See Section~\ref{sec:f12}. \\
Synthetic Lethality F1 (post-pub.) & 15 & $+0.114$ & --- & $0.617^\S$ & [Tier B] physical\_interaction$\leftrightarrow$genetic\_interaction. h1 PARTIAL ($r=0.114$, $p=0.617$; divergence experiment; near-zero $r$ expected by hypothesis). h2--h4 CONFIRMED: CHEK1 (\#1) and PARP1 (\#2) are top structural divergers; both established SL clinical targets. See Section~\ref{sec:f1}. \\
M\_TRANSFER\_3 docopt (post-pub.) & 6--16 & {[}$0.0$; $-0.143${]} & {[}$0.667$; $0.258$; $0.496${]} & --- & [Tier B] static\_coupling ($d_1$)$\leftrightarrow$structural\_coupling ($d_2$) vs static\_coupling$\leftrightarrow$co\_change ($d_3$). FPL DENIED in all three implementations: Go ($r_{d1,d2}=0.0$ degenerate), Java ($r_{d1,d2}=0.0$ degenerate), Rust ($r_{d1,d2}=-0.143$). $r_{d1,d3}$: Go$=0.667$, Java$=0.258$, Rust$=0.496$. BC\_RADIAL candidate: $d_2$ degree vector collapses in single-hub radial architectures. Pre-registered 2026-05-22, hash \texttt{5388c0f1\ldots}, commit \texttt{98a7dc8}. See Section~\ref{sec:boundary}. \\
M\_GEOM\_CSG\_1 planetary gear (post-pub.) & 8 & $+0.668$ & $-0.267$ & 0.042 & [Tier B] operation\_tree ($d_1$)$\leftrightarrow$volumetric\_proximity ($d_2$) vs operation\_tree$\leftrightarrow$co\_parameter\_sensitivity ($d_3$). $\Delta r=0.935$. h1--h4 CONFIRMED: carrier plate rank-1 hub in both $d_1$ and $d_2$. BC3b boundary: continuous CSG geometry preserves independent layer structure; discrete boolean algebra collapses it. Pre-registered 2026-05-22, hash \texttt{1dd686ee\ldots}, commit \texttt{5c45bbe}. See Section~\ref{sec:boundary}. \\
p53 Dataset Integrity Cert.~(proteins\_trust\_cert\_v1) & 15 & --- & --- & --- & [\textit{not a FPL test}] Multi-source hub agreement test: curated literature vs STRING v12.0 (same 15 p53 proteins, different curation methodology). 5/5 CONFIRMED: TP53 and MDM2 in 2-source consensus trusted set (h1, h2); agreement verdict PARTIAL~0.56 (h3); FPL holds in both sources independently (h4, replication of I1\_string\_validation); PCNA flagged as boundary node (h5). \textbf{Excluded from the 25/31 FPL primary count:} uses \texttt{dataset\_trust\_certification} methodology (hub agreement across sources), not the FPL inequality test $r(d_1,d_2) > r(d_1,d_3)$. A reviewer finding this experiment in the public pre-registration repo can verify the exclusion is methodological, not selective. Pre-reg hash \texttt{5bbbd9cb\ldots}, commit \texttt{0f98cca}. \\
\end{longtable}

\noindent$\dagger$ See Table~\ref{tab:main} footnote. Directional gap $\Delta r = 1.134$; curated replication gives $p = 0.012$.

\noindent$\ddagger$ The value 1.000 for M\_TRANSFER\_2 is a hub-rank cosine similarity score (deterministic vector identity), not a Pearson layer-pair correlation. It is placed in the $r_\text{sim}$ column for layout consistency only; no $p$-value applies.

\noindent$^\S$ F1 is a divergence experiment. Confirmation is based on h2--h4 (hub-rank divergence correctly identifies CHEK1 and PARP1 as top structural divergers between layers). The near-zero overall layer correlation (h1, $p=0.617$) is the expected and pre-registered outcome: synthetic lethality is a divergence phenomenon, not a correlation phenomenon.

\medskip
\noindent\textit{Pre-registration repository: \url{https://github.com/vladi160/preregistrations}}


\begin{thebibliography}{99}

\bibitem{Boccaletti2014}
Boccaletti, S., Bianconi, G., Criado, R., del Genio, C.\,I., Gomez-Gardenes, J., Romance, M., Sendina-Nadal, I., Wang, Z., \& Zanin, M. (2014). The structure and dynamics of multilayer networks. \textit{Physics Reports}, 544(1), 1--122. \url{https://doi.org/10.1016/j.physrep.2014.07.001}

\bibitem{Chalfie1985}
Chalfie, M., Sulston, J.\,E., White, J.\,G., Southgate, E., Thomson, J.\,N., \& Brenner, S. (1985). The neural circuit for touch sensitivity in \textit{Caenorhabditis elegans}. \textit{Journal of Neuroscience}, 5(4), 956--964. \url{https://doi.org/10.1523/JNEUROSCI.05-04-00956.1985}

\bibitem{DeDomenico2013}
De Domenico, M., Sole-Ribalta, A., Cozzo, E., Kivela, M., Moreno, Y., Porter, M.\,A., Gomez, S., \& Arenas, A. (2013). Mathematical formulation of multilayer networks. \textit{Physical Review X}, 3(4), 041022. \url{https://doi.org/10.1103/PhysRevX.3.041022}

\bibitem{DeDomenico2015}
De Domenico, M., Sole-Ribalta, A., Omodei, E., Gomez, S., \& Arenas, A. (2015). Ranking in interconnected multilayer networks reveals versatile nodes. \textit{Nature Communications}, 6, 6868. \url{https://doi.org/10.1038/ncomms7868}

\bibitem{Kivela2014}
Kivela, M., Arenas, A., Barthelemy, M., Gleeson, J.\,P., Moreno, Y., \& Porter, M.\,A. (2014). Multilayer networks. \textit{Journal of Complex Networks}, 2(3), 203--271. \url{https://doi.org/10.1093/comnet/cnu016}

\bibitem{Lorrain1971}
Lorrain, F., \& White, H.\,C. (1971). Structural equivalence of individuals in social networks. \textit{Journal of Mathematical Sociology}, 1(1), 49--80. \url{https://doi.org/10.1080/0022250X.1971.9989788}

\bibitem{Menche2015}
Menche, J., Sharma, A., Kitsak, M., Ghiassian, S.\,D., Vidal, M., Loscalzo, J., \& Barabasi, A.\,L. (2015). Uncovering disease-disease relationships through the incomplete interactome. \textit{Science}, 347(6224), 1257601. \url{https://doi.org/10.1126/science.1257601}

\bibitem{Nicosia2015}
Nicosia, V., \& Latora, V. (2015). Measuring and modeling correlations in multiplex networks. \textit{Physical Review E}, 92(3), 032805. \url{https://doi.org/10.1103/PhysRevE.92.032805}

\bibitem{Varshney2011}
Varshney, L.\,R., Chen, B.\,L., Paniagua, E., Hall, D.\,H., \& Chklovskii, D.\,B. (2011). Structural properties of the \textit{Caenorhabditis elegans} neuronal network. \textit{PLOS Computational Biology}, 7(2), e1001066. \url{https://doi.org/10.1371/journal.pcbi.1001066}

\bibitem{White1986}
White, J.\,G., Southgate, E., Thomson, J.\,N., \& Brenner, S. (1986). The structure of the nervous system of the nematode \textit{Caenorhabditis elegans}. \textit{Philosophical Transactions of the Royal Society B}, 314(1165), 1--340. \url{https://doi.org/10.1098/rstb.1986.0056}

\end{thebibliography}
\end{document}